\documentclass{jot}

\usepackage[utf8]{inputenc}
\usepackage[T1]{fontenc}
\usepackage[english]{babel}
\usepackage{microtype} 
\usepackage{tabularx} 
\usepackage{booktabs} 

\usepackage{proof}
\usepackage{amssymb}
\usepackage{alltt}
\usepackage{algorithm}

\newcommand {\pt}{{pt}}
\newcommand {\cpt}{{cpt}}
\newcommand {\copyn}{copy_n}
\newcommand {\Q}{{Q}}
\newcommand {\colors}{{colors}}
\newcommand {\equal}{{equal}}
\newcommand{\ObjPS}{$\lambda{{\cal O}bj^\oplus_S}$}

\newcommand {\pdg}[1]{\textcolor{black}{#1}}

\newcommand{\of}{{:}}

\def\frame{%
\medskip
\setbox0=\vbox\bgroup%
\advance\hsize by -0\fboxsep\advance\hsize by -2\fboxrule}

\def\endframe{%
\egroup\noindent\framebox[20pc][c]{\box0}\medskip}

\def\frametwo{%
\setbox0=\vbox\bgroup%
\advance\hsize by -0\fboxsep\advance\hsize by -2\fboxrule}

\def\endframetwo{%
\egroup\noindent\framebox[42pc][c]{\box0}}

\newcounter{pippo}

\def \msk {\hspace*{0.25em}}
\def \ssk {\hspace*{0.15em}}

\def \arrow {{\rightarrow}} 
\def \bredr {\mbox{${\rightarrow}_{\beta}$}}

\def \dbredr {\msk\arrow \kern-.75em\bredr\msk}
\def \dbreds {\arrow \kern-.75em\bredr}
\def \dbetared {\msk\arrow \kern-.75em\bredr\msk}
\def \dbredone {\mbox {${\rightarrow}\kern-.7em{\rightarrow}_{\beta}^1$}}

\newcommand{\range}{{\,i \in I}}

\newcommand{\rangej}{{j \in J}}
\newcommand{\rangeh}{{h \in H}}
\newcommand{\rangek}{{k \in K}}
\newcommand{\rangev}{{\,v \in V}}
\newcommand{\rangevi}{{\,v \in V_i}}
\newcommand{\rangewj}{{\,w \in W_j}}
\newcommand{\rangew}{{w \in W}}
\newcommand{\rangez}{{z \in Z}}

\newcommand{\newtypeIKJ}{\object{\m_i:{\upsilon_i}\sigma_i \diamond \m_j:{\upsilon_j}\sigma_j}^{\,i \in I{\cup}K}_\rangej}
\newcommand{\newtypeIJK}{\object{\m_i:{\upsilon_i}\sigma_i\diamond \m_j: {\upsilon_j}\sigma_j}^\range_{j \in J{\cup}K}}
\newcommand{\newtypeshift}{\object{\m_i:{\upsilon_i}\sigma_i \diamond \m_j:{\upsilon_j}\sigma_j}^{\,i \in I {\cup} \{k\}}_{j \in J{\setminus}\{k\}}}
\newcommand{\fnewtypeshift}{\ob{\m_i:\sigma_i \diamond \m_j:\sigma_j}^{\,i \in I {\cup} \{k\}}_{j \in J{\backslash}\{k\}}}
\newcommand{\newtypeshiftvw}{\object{\m_v:{\upsilon_v}\sigma_v \diamond \m_w:{\upsilon_w}\sigma_w}^{\,v \in V {\cup} \{j\}}_{w \in W{\setminus}\{j\}}}
\newcommand{\newtypevw}{\object{\m_v:{\upsilon_v}\sigma_v \diamond \m_w:{\upsilon_w}\sigma_w}^\rangev_\rangew}
\newcommand{\fnewtypevw}{\ob{\m_v:\sigma_v \diamond \m_w:\sigma_w}^\rangev_\rangew}
\newcommand{\fnewtypehk}{\ob{\m_h:{\upsilon_h}\sigma_h \diamond \m_k:{\upsilon_k}\sigma_k}^\rangeh_\rangek}
\newcommand{\newtypehk}{\object{\m_h:{\upsilon_h}\sigma_h \diamond \m_k:{\upsilon_k}\sigma_k}^\rangeh_\rangek}
\newcommand{\newtype}{\object{\m_i:{\upsilon_i}\sigma_i \diamond \m_j:{\upsilon_j}\sigma_j}^\range_\rangej}
\newcommand{\newtypeb}{\object{\m_i:{\upsilon'_i}\sigma'_i \diamond \m_j:{\upsilon'_j}\sigma'_j}^\range_\rangej}
\newcommand{\fnewtype}{\ob{\m_i:{\upsilon_i}\sigma_i \diamond \m_j:{\upsilon_j}\sigma_j}^\range_\rangej}
\newcommand{\ftype}{\ob{\m_i:{\upsilon_i}\sigma_i}^\range}
\newcommand{\stype}{\object{\m_i:{\upsilon_i}\sigma_i}^\range}
\newcommand{\stypej}{\object{\m_j:{\upsilon_j}\sigma_j}^\rangej}
\newcommand{\stypeb}{\object{\m_i:{\upsilon'_i}\sigma'_i}^\range}
\newcommand{\sbtype}{\object{\m_j:\tau_j}^\rangej}
\newcommand{\stypek}{\object{\m_k:{\upsilon_k}\sigma_k\{t\}}}
\newcommand{\stypev}{\object{\m_v:{\upsilon_v}\sigma_v\{t\}}^\rangev}
\newcommand{\ftypez}{\ob{\m_z:\sigma_z \diamond}^\rangez}
\newcommand{\ftypezk}{\ob{\m_z:\sigma_z \diamond}^{z \in Z{\cup}\{k\}}}
\newcommand{\ftypeh}{\ob{\m_h:\sigma_h \diamond}^\rangeh}
\newcommand{\fnewtypediamond} {\ob{\m_i:\sigma_i ~\diamond}^\range}
\newcommand{\newtypediamond} {\object{\m_i:{\upsilon}\sigma_i ~\diamond}^\range}
\newcommand{\newtypeomega} {\object{\m_i:\sigma_i \diamond \omega}^\range}
\newcommand{\fnewtypeomegavi} {\ob{\m_v:\sigma_v \diamond \omega}^\rangevi}
\newcommand{\fnewtypeomegavii} {\ob{\m_v:\sigma_v}^{v \in V_i{\cup}\{i\}}}
\newcommand{\fnewtypeomegahii} {\ob{\m_h:{\upsilon_h}\sigma_h}^{h \in H_i{\cup}\{i\}}}
\newcommand{\newtypeomegahii} {\object{\m_h:{\upsilon_h}\sigma_h}^{h \in H_i{\cup}\{i\}}}
\newcommand{\fnewtypeomegahi} {\ob{\m_h:\sigma_h}^{h \in H_i}}
\newcommand{\fnewtypeomegawjj} {\ob{\m_w:\sigma_w}^{w \in W_j{\cup}\{j\}}}
\newcommand{\fnewtypeomegakjj} {\ob{\m_k:\sigma_k}^{k \in K_j{\cup}\{j\}}}
\newcommand{\newtypeomegakjj} {\object{\m_k:\sigma_k}^{k \in K_j{\cup}\{j\}}}
\newcommand{\fnewtypeomegakj} {\ob{\m_k:\sigma_k}^{k \in K_j}}
\newcommand{\fnewtypeomegawj} {\ob{\m_w:\sigma_w \diamond \omega}^\rangewj}
\newcommand{\fnewtypeomega} {\ob{\m_v:\sigma_v \diamond \omega}^\rangev}
\newcommand{\fnewtypeomegai} {\ob{\m_i:\sigma_i \diamond \omega}^\range}
\newcommand{\fnewtypeomegaij} {\ob{\m_i:\sigma_i \diamond \omega}^{i \in I{\cup}J}}
\newcommand{\fnewtypeomegavk} {\ob{\m_v:\sigma_v}^{v \in V{\cup}\{k\}}}
\newcommand{\fnewtypeomegahk} {\ob{\m_h:\sigma_h}^{h \in H{\cup}\{k\}}}
\newcommand{\newtypeomegahk} {\object{\m_h:{\upsilon_h}\sigma_h}^{h \in H{\cup}\{k\}}}
\newcommand{\newtypeomegahj} {\object{\m_h:{\upsilon_h}\sigma_h}^{h \in H{\cup}\{j\}}}
\newcommand{\fnewtypeomegah} {\ob{\m_h:\sigma_h}^{h \in H}}
\newcommand{\fnewtypeomegavw} {\ob{\m_v:\sigma_v}^{v \in V{\cup}\{w\}}}
\newcommand{\fnewtypeb}{\ob{\m_i:\sigma_i \diamond \m_j:\sigma_j}}
\newcommand{\newtypec}{\object{\m_i:\sigma_i \diamond \m_j:\sigma_j}^\range_\rangej}
\newcommand{\fnewtypec}{\ob{\m_i:\sigma_i \diamond \m_j:\sigma_j}^\range_\rangej}
\newcommand{\abjtype}{\object{\m_i:{\upsilon_i}\sigma_i ~\diamond}^\range}
\newcommand{\fabjtype}{\ob{\m_i:\sigma_i ~\diamond}^\range}
\newcommand{\abjtypez}{\object{\m_z:\sigma_z}^\rangez}
\newcommand{\abjtypev}{\object{\m_v:{\upsilon_v}\sigma_v}^\rangev}
\newcommand{\abjtypeb}{\object{\m_i:\sigma_i}}

\newcommand {\match} {{\,\prec\!\!\!\sharp\,}}   

\newcommand{\deval} {\mbox{$\msk\arrow \kern-1.1em\eval\msk$}}

\newcommand{\eval} {\rightarrow} 
\newcommand{\ddeval} {\stackrel{det}{{\rightarrow}}}

\newcommand{\dlongeval}
 {\mbox{$\msk\longrightarrow \kern-1.1em\rightarrow\msk$}}

\newcommand{\devalu}{\mbox{$\msk\arrow \kern-1.1em\eval_u\msk$}}
\newcommand{\devalt}{\mbox{$\msk\arrow \kern-1.1em\eval_t\msk$}}

\newcommand{\ovatype}{\oplus}

\newcommand {\call} {{\,\Leftarrow\,}}

\newcommand {\add} {\leftarrow\hspace{-1em}+}
\newcommand {\ov} {\leftarrow}
\newcommand {\ova} {{\leftarrow\hspace{-0.75em}\oplus\,}}

\newcommand{\tr}{\ovatype}

\renewcommand {\v}{{v}}

\newcommand {\e}{e}

\newcommand {\bool}{{bool}}
\renewcommand {\int}{{int}}

\newcommand {\x}{{x}}
\newcommand {\f}{{f}}
\newcommand {\setx}{{set}_\x}
\newcommand {\addsetc}{{add_{col}}}

\newcommand {\g}{{g}}

\newcommand {\eq}{{eq}}

\newcommand {\col}{{col}}

\newcommand {\m}{{m}}

\renewcommand {\i}{{i}}

\newcommand {\p}{{p}}

\newcommand {\cp}{{cp}}

\newcommand {\q}{{q}}

\newcommand {\n}{{n}}

\newcommand {\getf}{{get_{\!\f}}}

\newcommand {\wrong}{{wrong}}

\newcommand {\extend}{{sel\!f_{ext}}}
\newcommand {\flyextend}{{fly_{ext}}}
\newcommand {\twoextend}{{inner_{ext}}}

\newcommand {\self}{{s\/}}

\newcommand {\addn}{{add_n}}
\newcommand {\addmn}{{add_{mn}}}
\newcommand {\addc}{{add_{col}}}

\renewcommand {\P}{{P}}
\newcommand {\CP}{{CP}}

\newcommand {\PClass}{{P_{class}}}

\newcommand {\new}{{new}}

\newcommand {\T}{\ast}
\newcommand {\TRIG}{\T_{rgd}}

\newcommand {\meth}[1]{\vec{#1}}

\newcommand {\vm}[1]{\vec{m}_{#1}}
\newcommand {\vn}[1]{\vec{n}_{#1}}
\newcommand {\vp}[1]{\vec{p}_{#1}}

\newcommand {\vs}[1]{\vec{\sigma}_{#1}}
\newcommand {\vsa}[1]{\vec{\sigma}_#1}

\newcommand {\class}[1]{class\ssk t.\l#1\r}
\newcommand {\clas}{class\ssk t.}

\newcommand {\pro}{pro\ssk t.}
\newcommand {\proa}{pro\ssk t'.}

\newcommand {\probject}[1] {pro\ssk t.\l#1\r}

\newcommand {\proj}[2]{pro\ssk t.#1\ovatype\vec{#2}}
\newcommand {\probjecu}[1] {pro\ssk t'.\l#1\r}

\newcommand {\obj}{obj\ssk t.}
\newcommand {\object}[1]{obj\ssk t.\l#1\r}

\newcommand {\objj}[2]{obj\ssk t.#1\ovatype\vec{#2}}

\newcommand {\ob}[1] {\l#1\r}

\def\vec#1{\overline{#1}}

\renewcommand {\l}{\langle}
\renewcommand {\r}{\rangle}

\newcommand {\ra}{\rangle}

\def \eqdef {\,\,\mbox{\small $\stackrel{\mbox{\tiny $\triangle$}}{=}$}\,\,}
\newcommand{\comment}[1]{}

\newcommand{\Obj}{$\lambda{{\cal O}bj}$}
\newcommand{\ObjP}{$\lambda{{\cal O}bj^\oplus}$}

\newcommand{\text}[1] {{\, \tr \langle #1 \rangle}}

\def\squareforqed{\hbox{\rlap{$\sqcap$}$\sqcup$}}
\def\qed{\ifmmode\squareforqed\else{\unskip\nobreak\hfil
\penalty50\hskip1em\null\nobreak\hfil\squareforqed
\parfillskip=0pt\finalhyphendemerits=0\endgraf}\fi}

\newtheorem{theorem}{Theorem}[section]
\newtheorem{corollary}[theorem]{Corollary}
\newtheorem{definition}[theorem]{Definition}
\newtheorem{example}{\it Example}[section]

\newtheorem{lemma}[theorem]{Lemma}

\newtheorem{proposition}[theorem]{Proposition}

\newtheorem{remark}[theorem]{Remark}

\newenvironment{proof-of}[1]{{\em Proof of #1:}}{}

\newif\ifeurasia
\eurasiafalse
\ifeurasia
\fi
\title{A prototype-based approach\\ to object reclassification}
\runningtitle{A protoype-based approach to object reclassification}

\author[affiliation=UniUD]{Alberto Ciaffaglione}{}
\author[affiliation=UniUD]{Pietro Di Gianantonio}{}
\author[affiliation=UniUD]{Furio Honsell}{}
\author[affiliation=INRIA]{Luigi Liquori}{}

\affiliation{UniUD}{DIMI, University of Udine, Italy}
\affiliation{INRIA}{Universit\'{e} C\^{o}te d'Azur, Inria, France}

\runningauthor{Ciaffaglione, Di Gianantonio, Honsell, Liquori}

\jotdetails{
}

\begin{document}

\begin{abstract}
We investigate, in the context of {\em functional prototype-based languages}, a calculus of objects which might extend themselves upon receiving a message, a capability referred to by Cardelli as a {\em self-inflicted} operation.
%
%
We present a sound type system for this calculus which guarantees that evaluating a
well-typed expression will never yield a \emph{message-not-found} runtime error.
%
%
The resulting calculus is an attempt towards the definition of a language combining the safety advantage of static type checking with the flexibility normally found in dynamically typed languages. 


\end{abstract}


\section{Introduction}\label{intro}



Object calculi and languages can be divided in the two main categories 
of class-based and prototype-based (a.k.a. object-based) ones. 
The latter, whose best-known example is JavaScript, provide the 
programmer with a greater flexibility compared to those classed-based, 
e.g. the possibility of changing at runtime the behaviour of objects, by 
modifying or adding methods.
Although such a flexibility is normally payed by the lack of static type systems, 
this is not necessarily the case, as it is possible to define a statically typed, prototype-based language.
One example in this direction was the {\em Lambda Calculus of Objects} (\Obj), 
introduced by Fisher, Honsell, and Mitchell \cite{FHM-94} as a first solid foundation 
for the prototyped-based paradigm.

\Obj\ is a lambda calculus extended with object primitives, where a new object 
may be created by modifying or extending an existing {\em
prototype}. The new object thereby inherits properties from the
original one in a controlled manner. Objects can be viewed as lists of
pairs {\em (method name, method body)} where the method body is (or
reduces to) a lambda abstraction whose first formal parameter is
always the object itself ({\tt this} in C$^{++}$ and Java).
The type assignment system of \Obj\ is set up so as to prevent the
unfortunate \emph{message-not-found} runtime error. Types of methods
are allowed to be specialized to the type of the inheriting
objects. This feature is usually referred to as ``mytype method
specialization''. The high mutability of method bodies is
accommodated in the type system via an implicit form of {\em higher-order polymorphism}, 
inspired by the the work of Wand on extensible records \cite{Wand-87}.

The calculus \Obj\ spurred an intense research in type assignment
systems for object calculi. Several calculi inspired by \Obj,
dealing with various extra features such as incomplete objects,
subtyping, encapsulation, imperative features, have appeared soon afterwards (see 
e.g. \cite{FM-95,BL-95,
BBDL-97,FM-98,BF-98}).



More specifically, \Obj\ supports two operations which may change the
shape of an object: method {\em addition} and method {\em override}.
The operational semantics of the calculus allows method bodies in objects 
to modify their own self, a powerful capability referred to by Cardelli as a 
{\em self-inflicted} operation \cite{Cardelli-95}.
%
%

Consider the method $\setx$ belonging, among others, to a $\pt$ object with an
$\x$ field:
$$
\pt \eqdef \l 
x{=} \lambda \self.0,\
\setx{=}\lambda \self.\lambda v. \l s \ov x{=}\lambda \self '.v \r,\
 \ldots \r
$$
%
%
\noindent When $\setx$ is called to $\pt$ with argument ``$3$'', written as $\pt \call \setx (3)$, 
the result is a new object 
where the $\x$ field has been set (i.e. overridden) to $3$.
Notice the self-inflicted operation of object override (i.e.
$\ov$) performed by the $\setx$ method.

However, in all the type systems for calculi of objects, both
those derived from \Obj\ and those derived from Abadi and Cardelli's 
foundational \emph{Object Calculus} \cite{AC-book},
the type system prevents the possibility for a method to self-inflict
an extension to the host object. We feel that this is an unpleasant
limitation if the message-passing paradigm is to be taken in
full generality. Moreover, in \Obj\ this limitation appears arbitrary,
given that the operational semantics supports without difficulty
self-inflicted extension methods.

There are plenty of situations, both in programming and in real life,
where it would be convenient to have objects which modify their
interface upon an execution of a message. Consider for instance the
following situations.

\begin{itemize}
\item The process of {\em learning} could be easily modeled using an
 object which can react to the ``teacher's message'' by extending its
 capability of performing, in the future, a new task in response to a
 new request from the environment (an old dog could appear to learn
 new tricks if in his youth it had been taught a ``self-extension''
 trick).
 
\item The process of ``vaccination'' against the virus ${\cal X}$ can
 be viewed as the act of extending the capability of the immune
 system of producing, in the future, a new kind of ``${\cal
 X}$-antibodies'' upon receiving the message that an ${\cal
 X}$-infection is in progress.
Similar processes arise in epigenetics.

\item In standard typed class-based languages the structure of
 a class can be modified only statically.
 If we need to add a new method to an instance
of a class we are forced to recompile
 the class and to make the modification needlessly
 available to all the class instances, thereby wasting
 memory. If a class had a self-extension method, only the instances
 of the class which have dynamically executed this method would
 allocate new memory, without the need of any re-compilation.
As a consequence, many sub-class declarations could be easily explained away if
 suitable self-exten\-sion methods in the parent class were available.
 

\item {\em Downcasting} could be smoothly implementable on objects
 with self-extension methods. For example, for a colored point $\cpt$ extending the $\pt$ object above, the following expression could be made to type check (details in Section \ref{examples}):
$${\cpt \call \eq(\pt \call \addsetc \, (black))}$$
where $\addsetc$ is intended to be a self-extension method of $\pt$ (adding a $\col$ method)
and $\eq$ is the name of the standard binary equality method.

\item Self-extension is strictly related to object evolution 
and object reclassification (see Sections \ref{sec:reclass} and \ref{related}), 
two features which are required in areas such as e.g. banking, GUI development, and games.

\end{itemize}
Actually, the possibility of modifying objects at runtime is already available in 
dynamically typed languages such as Smalltalk 
(via the \texttt{become} method), Python (by modifying the \texttt{\_class\_} attribute), and Ruby.
On the other hand, the self-extension itself is present, and used, in the prototype-based JavaScript language.


In such a scenario, the goal of this paper is to introduce the prototype-based 
\ObjP, a lambda calculus of objects in the style of \Obj, together with a type
assignment system which allows self-inflicted
extension still catching statically the message-not-found
runtime error. This system can be further extended to accommodate
other subtyping features; by way of example we will present a
``width-subtyping" relation that permits sound method override and a
limited form of object extension.
In fact, this manuscript completes and extends the paper \cite{DBLP:conf/oopsla/GianantonioHL98}.

We remark that the research presented in this article belongs to a series of similar investigations 
\cite{Zhao10,ChughHJ12,Zhao12}, whose aim is to define more and more powerful
type assignment systems, capable to statically type check larger and larger fragments 
of a prototype-based, dynamically typed language like JavaScript.
The ultimate goal is the definition of a language combining the safety advantage of 
static type checking with the flexibility normally found in dynamically typed languages. 
 
\subsubsection*{Self-inflicted extension}
To enable the \ObjP\ calculus to perform self-inflicted extensions,
two modifications of the system in \cite{FHM-94} are necessary. The
first is, in effect, a simplification of the original syntax of the
language. The second is much more substantial and it involves the type
discipline.

As far as the syntax of the language is concerned, we are forced to
unify into a {\em single} operator, denoted by $\ova$, the two
original object operators of \Obj, i.e. object extension
($\add$) and object override ($\ov$). This is due to the fact that,
when iterating the execution of a self-extension method, only the
first time we have a genuine object extension, while from the second time on we have just a simple object override. 
\begin{example}\label{intro-ex}
Consider the $\addsetc$ method, that adds a $\col$ field to the ``point'' object $\p$:
%
$$
\p \eqdef \l 
x{=} \lambda \self.0,\
\setx{=}\lambda \self.\lambda v. \l s \ova x{=}\lambda \self '.v \r,\
\addsetc{=}\lambda \self.\lambda \v.\l \self \ova \col {=} \lambda \self'.\v \r
\r
$$ 
%
 When $\addsetc$ is sent to $\p$ with argument ``$white$'', 
 i.e. $\p \call \addsetc(white)$, the result is
 a new object $\cp$ where the $\col$ field has been added to $\p$
 and set to $white$:
$$
\cp \eqdef \l 
x{=} \ldots,\
\setx{=} \ldots,\
\addsetc{=} \ldots,\
\col {=} \lambda \self. white
\r
$$ 
If $\addsetc$ is sent twice to $\p$,
i.e. $\cp \call \addc(black)$,
then, since the $\col$ field is already present in $\cp$, 
it will be overridden with the new ``$black$'' value:
$$
\cp' \eqdef \l 
x{=} \ldots,\
\setx{=} \ldots,\
\addsetc{=} \ldots,\
\col {=} \lambda \self. white,\
\col {=} \lambda \self. black
\r
$$ 
Therefore, only the rightmost version of a method will be the effective one.
\end{example}


As far as types are concerned, we add two new kinds of object-types, namely
$\tau \ovatype \m$, which can be seen as the type theoretical counterpart of
the syntactic object $\l \e_1 \ova \m = \e_2 \r$,
and $\pro{R} \ovatype \m_1 \ldots \ovatype \m_k$,
a generalization of the original $\clas R$ in \cite{FHM-94}, named $pro$-type.
%
Intuitively, if the type $\pro R \ovatype \m_1 \ldots \ovatype \m_k$ 
is assigned to an object $\e$  ($t$ represents the type of self), $\e$ can respond to all the methods $\m_1, \ldots, \m_k$.
Mandatory, the list of pairs $R$ contains all the methods $\m_1, \ldots, \m_k$ together with their corresponding types; moreover, $R$ may contain some {\em reserved} methods, i.e. methods that can be added to $\e$ either by ordinary object-extension or by a 
method in $R$ which performs a self-inflicted extension (therefore, if $R$ did not contain reserved methods, $\pro R \ovatype \m_1 \ldots \ovatype \m_k$ would coincide with $\clas R$ of \cite{FHM-94}).

To convey to the reader the intended meaning of $pro$-types, let us suppose that an object $\e$ is assigned the type 
$\probject{\m\of t \ovatype \n, \n\of\int}\ovatype \m$.
In fact, $\e \call n$ is not typable, but as
$\e \call \m$ has the effect
of adding the method $\n$ to the interface of $\e$, thus updating
the type of $\e$ to $\probject{\m\of t \ovatype \n, \n\of int} \ovatype \m \ovatype \n$, then $(\e \call m) \call n$ is typable.
 
The list of reserved methods in a $pro$-type is crucial to
enforce the soundness of the type assignment system. Consider e.g. an 
object containing two methods, $addn_1$, and $addn_2$, each 
of them self-inflicting the extension of a new method $\n$.
The type assignment system has to carry enough information so as to enforce 
that the same type will be assigned to
$\n$ whatever self-inflicted extension has been executed.


The typing system that we will introduce ensures that we can always
dynamically add new fresh methods for $pro$-types, thus leaving
intact the original philosophy of rapid prototyping, peculiar to
object calculi.

To model specialization of inherited methods, we use the
notion of {\em matching}, a.k.a. type extension, originally introduced by
Bruce \cite{Bruce-94} and later applied to the Object Calculus
\cite{AC-book} and to \Obj\ \cite{BB-97}. At the price of a little
more mathematical overhead, we could have used also the implicit
higher-order polymorphism of \cite{FHM-94}.

\subsubsection*{Object subsumption.} 
As it is well-known, see e.g. \cite{AC-book,FM-94}, the
introduction of a subsumption relation over object-types makes the
type system unsound. In particular, width-subtyping clashes with
object extension, and depth-subtyping clashes with object override.
In fact, on $pro$-types no subtyping is possible. In order to
accommodate subtyping, we add another kind of object-type, 
i.e. $\obj R \ovatype \m_1 \ldots \ovatype \m_k$, which behaves like 
$\pro R \ovatype \m_1 \ldots \ovatype \m_k$ except that it can 
be assigned to objects which can be extended only by making 
longer the list $\ovatype \m_1 \ldots \ovatype \m_k$ (by means of reserved methods that appear in $R$). On $obj$-types a (covariant) width-subtyping is 
permitted\footnotemark. 


\footnotetext{The $pro$ and $obj$ terminology is the same as in 
 Fisher and Mitchell \cite{FM-95,FM-98}.}

\paragraph{Synopsis.} 
The present paper is organized as follows.
In Section \ref{calculus} we introduce the calculus \ObjP, its small-step
operational semantics, and some intuitive examples to
illustrate the idea of self-inflicted object extension.
In Section \ref{types} we define the type system for \ObjP\ and discuss in 
detail the intended meaning of the most interesting rules.
In Section \ref{subtyping} we show how our type system is compatible with
a width-subtyping relation.
Section \ref{examples} presents a collection of typing examples.
In Section \ref{sec:soundness} we state our soundness result, namely that every closed and well-typed expression will not produce wrong results.
Section \ref{sec:reclass} is devoted to workout an example, to illustrate the potential of the self-inflicted extension mechanism as a runtime feature, in connection with object reclassification.
In Section \ref{related} we discuss related work.
The complete set of type assignment rules appears in the Appendix, together
with full proofs.
%

The present work extends and completes \cite{DBLP:conf/oopsla/GianantonioHL98} in the following way: we have slightly changed the reduction semantics, substantially refined the type system, fully documented the proofs, and, in the last two novel sections, we have connected our approach with the related developments in the area.

 



\section{The lambda calculus of objects}\label{calculus}


In this section, we present the Lambda Calculus of Objects \ObjP. The
terms are defined by the following abstract grammar:
$$
\begin{array}{lclr}
 \e & ::= & c \mid x \mid 
 \lambda x.\e \mid 
 \e_1 \e_2 \mid 
 & ~~\mbox{($\lambda$-terms)}
 \\[3mm] 
 & &\l\r \mid 
 \l \e_1 \ova \m = \e_2 \r \mid
 \e \call \m \mid 
 & ~~\mbox{(object-terms)}
 \\[3mm] 
 & & Sel(\e_1,\m,\e_2) 
 & ~~\mbox{(auxiliary-terms)}
 \end{array}
$$
where $c$, $x$, $m$ are meta-variables ranging over sets of constants, variables, and names of methods, respectively.
As usual, terms that differ only in the names of bound variables are identified.
Terms are untyped $\lambda$-terms enriched with objects: the intended meaning 
of the object-terms is the following: $\l\,\r$ stands
for the empty object; $\l \e_1 \ova \m =\e_2 \r$ stands
for extending/overriding the object $\e_1$ with a method $\m$ whose
body is $\e_2$; $\e \call \m$ stands for the result of sending
the message $\m$ to the object $\e$.

The auxiliary operation $Sel(\e_1,\m,\e_2)$ searches the body
of the $\m$ method within the object $\e_1$.
In the recursive search of $\m$, $Sel(\e_1,\m,\e_2)$ removes methods from 
$\e_1$; for this reason we need to introduce the expression $\e_2$,
which denotes a function that, applied to $e_1$, reconstructs the original
object with the complete list of its methods.
This function is peculiar to the operational semantics and, in practice, 
could be made not available to the programmer.

To lighten up the notation, we write $\l \m_1{=}\e_1,\ldots, \m_k{=}\e_k \r$ as syntactic sugar for  $\l \ldots \l \l \r \ova \m_1{=}\e_1 \r \ldots \ova \m_k{=}\e_k\r$, where $k {\geq} 1$. 
Also, we write $e$ in place of $\lambda x.e$ if $x {\notin} FV(e)$; this mainly concerns methods, whose first formal parameter is always their host object: e.g. $\lambda s.1$ and $\lambda \self'.(\self \call m)$
are usually written $1$ and $\self \call m$,
respectively. 
 
\subsection{Operational semantics}
We define the semantics of \ObjP\ terms by means of the reduction
rules displayed in Figure \ref{small-step} (small-step semantics $\eval$);
the evaluation relation $\deval$ is then taken to be the symmetric, 
reflexive, transitive and contextual closure of $\eval$.
 
In addition to the standard $\beta$-rule for $\lambda$-calculus, the
main operation on objects is method invocation, whose reduction is
defined by the $(Selection)$ rule. Sending a message $\m$ to an object
$\e$ which contains a method $\m$ reduces to $Sel(\e,\m,\lambda s.s)$, where the
arguments of $Sel$ have the following intuitive meanings:
\begin{description}
\item[\rm $1^{st}$-arg.] is a sub-object of the receiver (or recipient) of the message;
\item[\rm $2^{nd}$-arg.] is the message we want to send to the receiver;
\item[\rm $3^{rd}$-arg.] is a function that transforms the first argument in the original receiver.
\end{description}
By looking at the last two rules, one may note that the
$Sel$ function scans the receiver of the message until it finds
the definition of the called method: when it finds
such a method, it applies its body to the receiver of the message.
Notice how the $Sel$ function carries over, in its search, all the informations 
necessary to reconstruct the original receiver of the message.
The following reduction illustrates the evaluation mechanism:
$$
\begin{array}{lcl}
\ob{id = \lambda s . s, \ one =  1} \call id \ & \eval
\\
Sel(\ob{id = \lambda s . s, \ one =  1}, \ id , \ \lambda s' . s') \ & \eval
\\
Sel(\ob{id = \lambda s . s}, \ id , \  \lambda s'' . (\lambda s' . s')\l s'' \ova one = 1 \r) \ & \eval
\\
Sel(\ob{id = \lambda s . s}, \ id , \  \lambda s'' . \l s'' \ova one = 1 \r) \ & \eval \
\\
(\lambda s . s) (( \lambda s'' . \l s'' \ova one = 1 \r) \ob{id = \lambda s . s}) \ &\eval \ & \ob{id = \lambda s . s, \ one =  1}
\end{array}
$$
That is,  in order to call the first method $id$ of an object-term with two methods,  $\ob{id = \lambda s . s, \ one =  1}$, one needs to consider the subterm containing just the first method  $\ob{id = \lambda s . s}$  and construct a function, $\lambda s'' . \l s'' \ova one = 1 \r  $, transforming the subterm in the original term.

\begin{figure}[t]
$$
 \begin{array}{llcl}
 (Beta) & (\lambda x.\e_1) \e_2 
 & \eval & \e_1[\e_2/x]
 \\[3mm] 
 (Selection) & \e \call \m
 & \eval & Sel(\e, \m, \lambda s.s) 
 \\[3mm] 
 (Success) & Sel(\l \e_1 \ova \m = \e_2 \r, \m, \e_3) 
 & \eval & \e_2 (\e_3 \l \e_1 \ova \m = \e_2 \r)
 \\[3mm] 
 (Next) & Sel(\l \e_1 \ova \n = \e_2 \r, \m, \e_3) 
 & \eval & Sel(\e_1, \m, \lambda s . e_3 \l s \ova \n = \e_2 \r)
 \end{array}
$$
\caption[]{Reduction Semantics (Small-Step)}
\label{small-step}
\end{figure}
 
\begin{proposition}
 The $\eval$ reduction is Church-Rosser.
\end{proposition}
A quite simple technique to prove the Church-Rosser property for the 
$\lambda$-calculus has been proposed by Takahashi \cite{Takahashi95}.
The technique is based on parallel reduction and on Takahashi translation.
It works as follows: first one defines a parallel reduction on $\lambda$-terms, 
where several redexes can be reduced in parallel; then one shows that for any 
term $\e$ there is a term $\e^*$, i.e. Takahashi's translation, obtained from $M$ 
by reducing a maximum set of redexes in parallel.
It follows almost immediately that the parallel reduction satisfies the 
triangular property, hence the diamond property, and therefore the calculus in confluent.
With respect to the $\lambda$-calculus, \ObjP\ contains, besides the $\lambda$-rule, reduction rules for object terms; however, the latter do not interfere with the former, hence Takahashi's technique can be applied to the \ObjP\ calculus.

A deterministic, call by name, evaluation strategy over terms $\ddeval$ may be defined on \ObjP\ by restricting 
the set of contexts used in the contextual closure of the reduction relation.
%
In detail, we restrict the contextual closure to the set of contexts generated by the following grammar:
$$
C[\ ] = [\ ] \mid C[\ ] \e \mid C[\ ] \call \m \mid Sel(C[\ ] , \m, \e) 
$$
%
The set of values, i.e. the terms that are well-formed (and typable according to the type system we introduce in Section \ref{types}) and where no reduction is possible, is defined by the following grammar:
$$
\begin{array}{lcl}
 obj & ::= & \l \r \mid \l \e_1 \ova \m = \e_2 \r \\[3mm]
 v & ::= & c \mid \lambda x.\e \mid obj 
\end{array}
$$
%
 
\subsection{Examples}\label{sub-examples}

 In the next examples we show three objects,
performing, respectively:
\begin{itemize}
\item a self-inflicted extension;
\item two (nested) self-inflicted extensions;
\item a self-inflicted extension ``on the fly''.
\end{itemize}

\begin{example}
\rm 
\label{extend} 
Consider the object $\extend$, defined as follows:
\begin{eqnarray*}
 \extend & \eqdef & 
 \l \addn = \lambda \self. 
 \l \self \ova \n{=}1 
 \r \r.
\end{eqnarray*}
If we send the message $\addn$ to $\extend$, then we have the
following computation:
\begin{eqnarray*}
 \extend \call \addn 
 & \eval & Sel(\extend,\addn,\lambda s'.s') \\
 & \deval & (\lambda \self. 
 \l \self \ova \n{=}1 \r) \extend\\
 & \eval & \l \extend \ova \n {=}1 \r
\end{eqnarray*}
i.e. the method $\n$ has been added to $\extend$. If we send the message $\addn$ twice to $\extend$, i.e. $\l \extend \ova \n {=} 1 \r \call \addn$, the
method $\n$ is only overridden with the same body; hence, we get an
object which is ``operationally equivalent'' to the previous one.

\end{example}

\begin{example}
\rm 
\label{twoextend} 
Consider the object $\twoextend$, defined as follows:
$$\begin{array}{l}
\twoextend \eqdef
\l \addmn = \lambda \self.
 \l \self \ova \m{=}
 \lambda s'. \l s' \ova \n{=}
 1 \r \r \r
\end{array}
$$
If we send the message $\addmn$ to $\twoextend$, then we obtain:
$$\begin{array}{c}
\twoextend \call \addmn 
\ \deval\ 
\l \twoextend \ova \m {=}\lambda s. \l s \ova \n {=}
 1 \r \r
\end{array}$$
i.e. the method $\m$ has been added to $\twoextend$. On the other
hand, if we send first the message $\addmn$ and then $\m$ to
$\twoextend$, both the methods $\m$ and $\n$ are added:
$$\ 
(\twoextend \call \addmn) \call \m 
\quad \deval \quad 
\begin{array}{lcl} 
\l \addmn & {=} & \lambda \self. \l \self \ova 
 \m {=} \lambda s'. \l s' \ova 
 \n {=} 1 \r \r, \\
 ~ \m & {=} & \lambda \self. \l \self \ova 
 \n {=} 1 \r,\\ 
 ~ \n & {=} & 1 \r
\end{array}$$
\end{example}

\begin{example}
\rm 
\label{flyextend} 
Consider the object $\flyextend$, defined as follows:
$$
\flyextend \eqdef \l
\f {=} \lambda \self. \lambda s'. s' \call \n,\
\getf {=} \lambda \self.(\self \call \f) \l \self \ova \n {=} 1 \r \r
$$
%
If we send the message $\getf$ to $\flyextend$, then we get the
following computation:
\begin{eqnarray*}
\flyextend \call \getf
 & \eval& Sel(\flyextend,\getf,\lambda s''. s'') \\
 &\eval & (\lambda \self. (\self \call \f) 
 \l \self \ova \n {=} 1 \r) \flyextend\\
 &\eval & (\flyextend \call \f) \l \flyextend \ova \n {=} 1 \r\\
 &\eval & Sel(\flyextend,\f,\lambda s''.s'') 
 \l \flyextend \ova \n {=} 1\r\\
 &\deval & (\lambda \self. \lambda s'. s' \call \n) \flyextend 
 \l \flyextend \ova \n {=} 1 \r\\
 &\deval & \l \flyextend \ova \n {=} 1 \r \call \n\\
 &\deval & 1
\end{eqnarray*}
i.e. the following steps are performed:
\begin{enumerate}
\item the method $\getf$ calls the method $\f$ with actual parameter
 the host object itself augmented with the $\n$ method;
 
\item the $\f$ method takes as input the host object augmented with
 the $\n$ method, and sends to this object the message $\n$, which
 simply returns the constant $1$. 
 \end{enumerate} %
\end{example}


\section{Type system}\label{types}


In this section, we introduce the syntax of types and we discuss
the most interesting type rules. For the sake of simplicity, we
prefer to first present the type system without the rules 
related with object subsumption (which will be discussed in Section
\ref{subtyping}).
The complete set of rules can be found in Appendices A and B.

\subsection{Types}
The type expressions are described by the following grammar:
$$
\begin{array}{lclr}
\sigma & ::= & \iota \mid \sigma \arrow \sigma \mid \tau
& \quad\textrm{(generic-types)}
\\[3mm]
\tau & ::= & t \mid \pro{R} \mid \tau \ovatype m
& \quad\textrm{(object-types)}
\\[3mm]
R & ::= & \l\r \mid \l R,\m\of\sigma\r
& \quad\textrm{(rows)}
\\[3mm]
\kappa & ::= & \T
& \quad\textrm{(kind of types)}
\end{array}
$$
In the rest of the article we will use $\sigma$ 
as meta-variable ranging over generic-types,
$\iota$ over constant types,
$\tau$ over object-types.
Moreover, $t$ is a type variable, $R$ a metavariable ranging over rows, i.e. unordered sets of  pairs {\em (method label, method type)}, $\m$ a
method label, and $\kappa$ a metavariable ranging over the unique kind
of types $\T$.

To ease the notation, we write 
$\l \ldots \l \l \r ,m_1 \of \sigma_1 \r \ldots, m_k \of \sigma_k \r$  as  
$\l\m_1\of\sigma_1, \ldots, m_k \of \sigma_k\r$
or $\l \vm{k}\of\vs{k} \r$
or else simply $\l \vm{}\of \vs{}\r$ in the case the subscripts can be omitted.
Similarly, we write either $\tau \ovatype \vm{k}$
or $\tau \ovatype \vm{}$
for $\tau\ovatype \m_1 \ldots \ovatype \m_k$,
and $\tau \ovatype \vm{},\n$
for $\tau\ovatype \m_1 \ldots \ovatype \m_k \ovatype \n$.
If $R \equiv \l \vm{}\of \vs{}\r$, then we denote $\vm{}$ by $\meth{R}$,
and we write $R_1\subseteq R_2$ if $R_1 \equiv \l \vm{}\of\vsa{1}{} \r$ and $R_2 \equiv \l \vm{}\of\vsa{1}{},\vn{}\of\vsa{2}{} \r$.
%

As in \cite{FHM-94}, we may consider object-types as a form of recursively-defined types.
Object-types in the form $\proj{R}{m}$ are named $pro$-types, where $pro$ 
is a binder for the type-variable $t$ representing ``self'' (we use $\alpha$-conversion of type-variables bound by $pro$).
%
The intended meaning of a $pro$-type
$
\probject{\vm{} \of\vs{}} \ovatype \vn{}
$
is the following:
\begin{itemize}
\item the methods in $\vm{}$ are the ones which are present in the $pro$-type;

\item the methods in $\vn{}$, being in fact a subset of those in $\vm{}$, are the methods that are \emph{available} and can be invoked
(it follows that the $pro$-type $\probject{\vm{} \of\vs{}} \ovatype \vm{}$
corresponds exactly to the object-type $\class{\vm{} \of\vs{}}$ in \cite{FHM-94});

\item the methods in $\vm{}$ that do not appear in $\vn{}$ are methods that cannot be invoked: they are just \emph{reserved}.
\end{itemize}

In the end, we can say that the operator ``$\ovatype$'' is used to make active and usable those methods that were previously just reserved in a $pro$-type; essentially, $\ovatype$ is the ``type counterpart'' of the operator on terms $\ova$.
In the following, it will turn out that we can extend an object $e$ with a new method $\m$ having type $\sigma$ only if it is possible to assign to $e$ an object-type of the form $\probject{R,\m\of\sigma} \ovatype \vn{},\m$; this reservation mechanism is crucial to guarantee the soundness of the type system.

\subsection{Contexts and judgments}
The contexts have the following form:
$$
\Gamma ::= \varepsilon \mid 
 \Gamma,x\of\sigma \mid 
 \Gamma, t \match \tau
$$
Our type assignment system uses judgments of the following shapes: 
$$
\Gamma \vdash ok \quad \qquad
 \Gamma \vdash \sigma : \T \quad \qquad
 \Gamma \vdash e : \sigma \quad \qquad
 \Gamma \vdash \tau_1 \match \tau_2
$$
The intended meaning of the first three judgments is standard:
well-formed contexts and types, and assignment of 
type $\sigma$ to term $\e$.
The intended meaning of $\Gamma \vdash \tau_1 \match
\tau_2$ is that $\tau_1$ is the type of a possible extension of an
object having type $\tau_2$. As in \cite{Bruce-94}, and in \cite{BL-95,BBL-96,BBDL-97,BB-97}, this
judgment formalizes the notion of {\em method-specialization} (or
{\em protocol-extension}), i.e. the capability to ``inherit''
the type of the methods of the prototype.

\subsection{Well formed context and types}
The type rules for well-formed contexts are quite standard. We
just remark that in the $(Cont{-}t)$ rule:
$$
\infer
 {\Gamma, t \match \proj{R}{m} \vdash ok} 
 {\Gamma \vdash \proj{R}{m} : \T & 
 t \not \in Dom(\Gamma)}
$$
we require that the object-types used to bind variables are not variable types themselves: this condition does not have any serious restriction, and has been set
in the type system in order to make simpler the proofs of its properties.

The $(Type{-}Pro)$ rule:
$$
\infer
{\Gamma \vdash \probject{R,\m\of\sigma}:\T} 
{\Gamma, t \match \pro{R} \vdash \sigma :\T & \m \not \in \meth{R}
 }
$$
asserts that the object-type $\probject{R,\m\of\sigma}$ is well-formed
if the object-type $\pro{R}$ is well-formed and the type
$\sigma$ is well-formed under the hypothesis that $t$ is an
object-type containing the methods in $\meth{R}$. Since $\sigma$
may contain a subexpression of the form $t \ovatype \n$, with $\n \in \meth{R}$, 
we need to introduce in the context the hypothesis $t \match
\pro{R}$ to prove that $t \ovatype \n$ is a well-formed type.
 
The $(Type{-}Extend)$ rule:
$$ 
\infer
 {\Gamma \vdash \tau \ovatype \vm{} : \T}
 {\Gamma \vdash \tau \match \pro{R}
 & 
 \vm{} \subseteq \meth{R}
 }
$$
asserts that in order to activate the methods $\vm{}$ in the
object-type $\tau$, the methods $\vm{}$ need to be present (reserved) in $\tau$.

\subsection{Matching rules} \label{matchingsect} 

\medskip \noindent The $(Match{-}Pro)$ rule:
$$
 \infer
 {\Gamma \vdash \pro R_1 \ovatype \vm{}
 \match \pro R_2 \ovatype \vn{}}
 {\Gamma \vdash \pro R_1 \ovatype \vm{}: \T
 & 
 \Gamma \vdash \pro R_2 \ovatype \vn{} : \T
 & 
 R_2 \subseteq R_1
 & 
 \vn{} \subseteq \vm{}
 }
$$
asserts that an object-type with more reserved and more available
methods specializes an object-type with less reserved and less
available methods.

The $(Match{-}Var)$ rule:
$$
 \infer
 {\Gamma_1, t \match \tau_1, \Gamma_2 \vdash t \ovatype \vm{}
 \match \tau_2} 
 {\Gamma_1, t \match \tau_1, \Gamma_2 \vdash \tau_1 \ovatype \vm{} \match \tau_2} 
$$
makes available the matching judgments present in the context.
It asserts that, if a context contains the hypothesis that a type
variable $t$ specializes a type $\tau_1$, and $\tau_1$ itself, incremented
with a set of methods $\vm{}$, specializes a type $\tau_2$, then, by
transitivity of the matching relation, $t$, incremented by the methods
in $\vm{}$, specializes $\tau_2$.

The $(Match{-}t)$ rule:
$$
 \infer
 {\Gamma \vdash t \ovatype \vm{} \match t \ovatype \vn{}} 
 {\Gamma \vdash t \ovatype \vm{} : \T 
 & 
 \vn{} \subseteq \vm{}
 } 
$$
concerns object-types built from the same type variable, simply
asserting that a type with more available methods specializes
a type with less available methods.

\subsection{Terms rules}\label{type-terms}
The type rules for $\lambda$-terms are self-explanatory and hence they
need no further justification. 
Concerning those for object terms, the $(Empty)$ rule assigns to an empty
object an empty $pro$-type, while
%
the $(Pre{-}Extend)$ rule:
$$
 \infer
 {\Gamma \vdash e: \probject{R_1, R_2} \ovatype \vm{}}
 {\Gamma \vdash e : \pro{R_1} \ovatype \vm{}
 &
 \Gamma \vdash \probject{R_1, R_2} \ovatype \vm{}: \T
 }
$$
asserts that an object $e$ having type $\pro{R_1} \ovatype \vm{}$ can be
considered also an object having type $\probject{R_1, R_2} \ovatype \vm{}$, i.e. with more reserved methods.
This rule has to be used in conjunction with the
$(Extend)$ one; it ensures that we can dynamically add fresh methods.
Notice that $(Pre{-}Extend)$ \emph{cannot} be applied when $e$ is a variable $\self$ representing self; in fact, as explained in the
Remark \ref{spiego} below, the type of $\self$ can only be a type
variable. This fact is crucial for the soundness of the type system.

The $(Extend)$ rule:
$$ 
 \infer
 {\Gamma \vdash 
 \langle e_1 \ova \n = e_2 \rangle : \tau \ovatype \n}
 {\begin{array}{l}
 \Gamma \vdash e_1 : \tau \qquad 
 \Gamma \vdash \tau \match 
 \probject{R, \n\of\sigma} \ovatype \vm{} \\[1mm]
 \Gamma, t \match \probject{R ,\n\of\sigma} \ovatype \vm{},\n \vdash
 \e_2 : t \arrow \sigma
 \end{array}
 }
$$
can be applied in the following cases:
\begin{enumerate}
\item when the object $\e_1$ has type $\probject{R, \n\of\sigma} \ovatype \vm{}$ 
 (or, by a previous application of the $(Pre{-}Extend)$ rule, 
 $\pro{R} \ovatype \vm{}$).
 In this case the object $\e_1$ is extended with the (fresh) method $\n$;
 
\item when $\tau$ is a type variable $t$. In this case $\e_1$ can be
 the variable \self, and a {\em self-inflicted extension} takes place.
\end{enumerate} 
The bound for $t$ is the same as the final type for the object $\l
\e_1 \ova \n = \e_2 \rangle$; 
\pdg{this allows a recursive call of the method $\n$ inside the expression 
$\e_2$, defining the method $\n$ itself.} 

The $(Override)$ rule:
$$
 \infer
 {\Gamma \vdash 
 \langle e_1 \ova \n = e_2 \rangle: \tau}
 {\begin{array}{l}
 \Gamma \vdash e_1 : \tau \qquad 
 \Gamma \vdash \tau \match 
 \probject{R, \n\of\sigma} \ovatype \vm{},\n \\[1mm]
 \Gamma, t \match \probject{R ,\n\of\sigma} \ovatype \vm{},\n \vdash
 \e_2 : t \arrow \sigma
 \end{array}
 }
$$
is quite similar to the $(Extend)$ rule, but it is applied 
when the method $\n$ is {\em already} available in the object $\e_1$, 
hence the body of $\n$ is {\em overridden} with a new one.

\begin{remark} \label{spiego}
 By inspecting the $(Extend)$ and $(Override)$ rules,
 one can see why the type of the object itself is always a type variable.
 In fact, the body $\e_2$ of the new added method $\n$ needs to have type 
 $t \arrow \sigma$. Therefore, if $\e_2$ reduces to a value, this value
 has to be a $\lambda$-abstraction in the form 
 $\lambda \self . e_2'$. It follows that, in assigning a type to $\e_2'$,
we must use a context containing the hypothesis $\self : t$.
Since no subsumption rule is available, the only type we can deduce for 
 $\self$ is $t$. \qed
\end{remark}

\noindent
The $(Send)$ rule:
$$
\infer
 {\Gamma \vdash e \call \n: \sigma[\tau/t]}
 {\Gamma \vdash e : \tau
 & 
 \Gamma \vdash \tau \match \probject{R,\n\of\sigma} \ovatype \vm{},\n
 }
$$
is the standard rule that one can expect from a type system based on
matching. We require that the method we are invoking is available
in the recipient of the message.

In the $(Select)$ rule:
$$
 \infer 
 {\Gamma \vdash Sel(\e_1,\n,\e_2) : \sigma[\tau \ovatype \vn{}/t]}
 {\begin{array}{l}
 \Gamma \vdash \e_1 : \tau \qquad 
 \Gamma \vdash \tau \match 
 \probject{R,\n\of\sigma} \ovatype \vm{},\n \\[1mm]
 \Gamma, t \match \probject{R,\n\of\sigma} \ovatype \vm{},\n \vdash 
 \e_2 : t \arrow t \ovatype \vn{}
 \end{array}
 }
$$
the first two conditions ensure that the $\n$ method is available
in $\e_1$, while the last one that $e_{2}$ is a function that transforms 
an object into a more refined one.


\section{Dealing with object subsumption}\label{subtyping}


While the type assignment system \ObjP, presented in Section \ref{types}, allows self-inflicted extension, it does not allow object \emph{subsumption}.
This is not surprising: in fact, we could (by subsumption) first hide a method in an object, and then add it again with a type incompatible with the previous one.
The papers \cite{AC-book,FM-94,FHM-94,BL-95}
propose different type systems for prototype-based languages, where subsumption is permitted only in absence of object extension (and a fortiori self-inflicted extension). 
In this section, we devise a conservative extension of \ObjP, that we name \ObjPS\ (Appendix B collects its extra rules), to accommodate width-subtyping.
%
%
%
%

In the perspective of adding a subsumption rule to the typing system, we introduce another kind of object-types, i.e. $\objj{R}{m}$, named $obj$-types.
The main difference between the $pro$-types and the $obj$-types consists 
in the fact that the $(Pre{-}Extend)$ rule cannot be applied when an object 
has type $\objj{R}{m}$; it follows that the type $\objj{R}{m}$ permits extensions 
of an object only by enriching the list $\vm{}$, i.e. by making active its reserved methods. 
This approach to subsumption is inspired by the one in \cite{FM-95,Liquori-Ecoop-97}. 
Formally, we need to extend the syntax of types by means of $obj$-types and the kind of \emph{rigid}, i.e. non-extensible, types:
$$
\begin{array}{lclr}
\tau & ::= & \ldots \mid \obj{R}
& \quad\textrm{(object-types)}
\\[3mm]
\kappa & ::= & \ldots \mid \TRIG
& \quad\textrm{(kind of types)}
\end{array}
$$
%
The subset of rigid types contains the $obj$-types and is closed under the arrow constructor. In order to axiomatize this, we introduce the judgment
$
\Gamma \vdash \tau : \TRIG
$,
whose rules are reported in Appendix \ref{appendix:sub}.
%
Intuitively, we can use the matching relation as a subtyping relation 
only when the type in the conclusion is rigid:
$$
\infer[(Subsume)]
{\Gamma \vdash e:\tau_2}
{\Gamma \vdash e: \tau_1 & \Gamma \vdash \tau_1 \match \tau_2 & 
& \Gamma \vdash \tau_2 : \TRIG}
$$
This is in fact is the rule performing object subsumption: it allows to use objects with an extended signature in any context expecting objects with a shorter one.

It is important to point out that, so doing, we do not need to introduce another partial order on types, 
i.e. an ordinary subtyping relation, to deal with subsumption. By
introducing the sub-kind of rigid types, we make the matching relation
compatible with subsumption, and hence we can make it play the role of
the width-subtyping relation. This is in sharp contrast with the uses of matching 
proposed in the literature (\cite{Bruce-94,BPF-97,BB-97}).
Hence, in our type assignment system, the matching is
a relation on types compatible with a limited subsumption rule.

Most of the rules for $obj$-types are a rephrasing of the rules presented so far, replacing the binder $pro$ with $obj$.
We remark that the $(Type{-}Obj{-}Rdg)$ rule
$$
 \infer 
 {\Gamma \vdash \objj{\l \vm{k}:\vs{k} \r}{n} : \TRIG}
 {\Gamma \vdash \objj{\l \vm{k}:\vs{k} \r}{n} : \T
 & \forall i \leq k. \ \Gamma \vdash \sigma_i : \TRIG
 \land t ~\mbox{covariant in}~\sigma_i}
$$
%
asserts that subsumption is unsound for methods having $t$ in
contravariant position with respect to the arrow type constructor.
Therefore, the variable $t$ is forced to occur only covariantly in $\vs{k}$.
A natural (and sound) consequence is that we cannot forget binary methods via subtyping (see \cite{BruEtAl96,Castagna-95,Castagna-book}).
%
%
The $(Promote)$ rule
$$
\infer 
 {\Gamma \vdash \proj{R_1}{m}
 \match \objj{R_2}{n}}
 {
 \Gamma \vdash \proj{R_1}{m}: \T
 & 
 \Gamma \vdash \objj{R_2}{n} : \T
 & 
 R_2 \subseteq R_1
 & 
 \vn{} \subseteq \vm{}
 }
$$
promotes a fully-specializable $pro$-type into a limitedly specializable $obj$-type with less reserved and less available methods. 

%







\section{Examples}\label{examples}


\begin{figure}[t]

Let be $\tau \eqdef \probject{\addn\of t \ovatype \n,\ \n\of\int}$ and $\Gamma \eqdef t \match \tau \ovatype \addn, s\of t$. Then:
$$
\infer[(Extend)]
{\vdash \l \addn = \lambda s. \l s \ova \n = 1\rangle\rangle:
 \tau \ovatype \addn}
 {\infer{\vdash \l\,\rangle: \tau}{\vdots} &
 \infer{\vdash \tau \match \tau}{\vdots}
 & \Delta}
 $$
 where the first two premises are derived straightforwardly and $\Delta$ as follows:
$$
\infer[(Abs)]
 {t \match \tau \ovatype \addn \vdash \lambda s.\l s \ova \n = 1\rangle:
 (t \to t \ovatype \n)}
 {\infer[(Extend)]
 {\Gamma \vdash \l s \ova \n = 1\rangle: t \ovatype \n}
 { \Gamma \vdash s: t
 & 
 \Gamma \vdash t \match \probject{\n \of \int}
 & 
 \Gamma, t' \match \probjecu{\n \of \int} \ovatype \n \vdash 1: t' \arrow \int
 }}
$$

%
\caption[]{A derivation for $\extend$}
\label{deriv}
\end{figure}

In this section, we give the types of the examples presented in
Section \ref{sub-examples}, together with some other motivating examples.
The objects $\extend$, $\twoextend$, and $\flyextend$, of Examples
\ref{extend}, \ref{twoextend}, and \ref{flyextend}, respectively, can be given the
following types:
\begin{eqnarray*}
\extend & : & \probject{\addn \of t \ovatype \n,\
\n \of \int} \ovatype \addn\\[2mm]
\twoextend & : & \probject{\addmn \of t \ovatype \m,\
 \m \of t \ovatype \n,\
 \n \of \int}\ovatype \addmn
 \\[2mm]
\flyextend & : & \probject{\f \of t \ovatype \n,\ 
 \getf \of t \ovatype \n \arrow \int,\
  \n \of \int} \ovatype \f, \getf
\end{eqnarray*}
A possible derivation for $\extend$ is presented in Figure \ref{deriv}.
%


\begin{example}\label{class-example}
We show how class declaration can be simulated in \ObjP\ and how using the self-inflicted extension we can factorize in a single declaration the definition of a hierarchy of classes.
Let the method $\addsetc$ be defined as in Example \ref{intro-ex}, and let us consider the simple class definition:
$$
\PClass \eqdef 
\l new = \lambda s . \l 
\n {=} 1,\
\addsetc {=} \lambda \self'.\lambda \x.\l \self' \ova \col {=} \x \r
\rangle \rangle
$$
%
Then, the object $\PClass$ can be used to create instances of both points and
colored points, by using the expressions:
$$
\PClass \call \new \qquad \mbox{and}
\qquad (\PClass \call \new)\call \addsetc(white)
$$
\end{example}

\begin{example}[Subsumption 1]\label{sub+ext}
We show how subsumption can interact with object extension. Let be: 
\begin{eqnarray*}
\P & \eqdef & \object {\n\of\int,\ \col \of \colors} \ovatype \n \\
\CP & \eqdef & \object{\n\of\int,\ \col \of \colors} \ovatype \n, \col \\
\g & \eqdef & \lambda s. \l s \ova \col = white \rangle
\end{eqnarray*}
and let $\p$ and $\cp$ be of type $\P$ and $\CP$, respectively. 
Then, we can derive:
$$
\begin{array}{l}
 \vdash \CP \match \P
\qquad
\vdash \g : \P \arrow \CP 
\qquad
\vdash \g (\cp): \CP 
\\
\vdash (\lambda \f. \equal(\f(\p)\call \col, \f(\cp)\call \col)) \g: \bool
\end{array}
$$ 
where the equality function $\equal$ has type $t \arrow t \arrow \bool$.
Notice that the terms:
$$
\g (\cp)
\qquad
(\lambda \f. \equal(\f(\p)\call \col,\f(\cp)\call \col))
$$
would not be typable without the subsumption rule.
\end{example}

\begin{example}[Subsumption 2]\label{nuovo}
We show how subsumption can interact with object self-inflicted extension.
Let be:
$$
\begin{array}{lcl}
\Q & \eqdef & \object{\n\of\int} \ovatype \n \\
\q & \eqdef &\l \copyn = \lambda \self.
 \lambda \self'. \l \self \ova \n = \self' \call \n \rangle\rangle
\end{array}
$$
By assuming $\p$ and $\cp$ as in Example \ref{sub+ext}, we can derive:
\begin{eqnarray*}
\vdash \q &:& \probject{\copyn\of \Q \to t \ovatype \n,\ \n\of\int}\ovatype \copyn 
\\ [2mm]
\vdash \q \call \copyn (\cp) &:& \probject{\n\of\int,\ \copyn\of \Q \to t} \ovatype \n, \copyn
\\ [2mm]
\vdash \q \call \copyn (\cp) \call \copyn (\p) &:& \probject{\n\of\int,\ \copyn\of \Q \to t} \ovatype \n, \copyn
\end{eqnarray*}
Notice in particular that the object
$\q \call \copyn (\cp) \call \copyn (\p)$ 
would not be typable without the subsumption rule.
\end{example}

\begin{example}[Downcasting]\label{downcasting}
The self-inflicted extension permits to perform explicit downcasting
simply by method calling. In fact,
let $\p_1$ and $\cp_1$ be objects with $\eq$ methods
(checking the values of $\n$ and the pairs $(\n,\col)$, respectively), and $\addsetc$ the self-extension method presented in Example \ref{class-example}, typable as follows:
$$
\vdash \p_1 : \pro R \quad~~~\mbox{and}~~~\quad
\vdash \cp_1 : \pro R \ovatype \col
$$
where $R \triangleq
 \ob{
 \n\of\int,\ 
 \eq\of t \arrow \bool,\ 
 \addsetc \of colors \to t \ovatype \col,\
 \col\of\colors
} 
 \ovatype \n, \eq, \addsetc
 $.
Then, the following judgments are derivable: 
\begin{eqnarray*}
\vdash \cp_1\call \eq &:& \pro R \ovatype \col \to \bool
\\
\vdash \p_1 \call \addsetc(white) &:& \pro R \ovatype \col
\\
\vdash \cp_1 \call \eq (\p_1 \call \addsetc(white)) &:& \bool
\end{eqnarray*}
\end{example}


\section{Soundness of the Type System}\label{sec:soundness}

In this section, we prove the crucial property of our type
system, i.e. the Subject Reduction theorem.
It needs a preliminary series of technical lemmas presenting basic
and technical properties, which are proved by inductive arguments.
As a corollary, we shall derive the fundamental result of the paper, i.e. the Type Soundness of our typing discipline.

We first address the plain type assignment system without subsumption \ObjP, then in Section \ref{sec:soundness2} we extend the Subject Reduction to the whole type system \ObjPS.
The proofs are fully documented in Appendices \ref{soundness} and \ref{soundness2}.

In the presentation of the formal results, we need $\alpha,\beta$ as metavariables for generic-types and $\rho, \upsilon$ for object-types.
Moreover, $\mathcal{A}$ is a metavariable ranging on statements in the forms $ok$, $\alpha : \T$, $\upsilon \match \rho$, $\e : \beta$, and $\mathcal{C}$ on statements in the forms $x\of\sigma$, $t \match \tau$.



\begin{lemma} (Sub-derivation) \label{sec:sub}
  \begin{itemize}
  \item[(i)] If $\Delta$ is a derivation of $\Gamma_1, \Gamma_2 \vdash
    \mathcal{A}$, then there exists a sub-derivation $\Delta' \subseteq \Delta$
    of $\Gamma_1 \vdash ok$.
  \item [(ii)] If $\Delta$ is a derivation of $\Gamma_1,
    x\of\sigma, \Gamma_2 \vdash \mathcal{A}$, then there exists a sub-derivation
    $\Delta' \subseteq \Delta$ of $\Gamma_1 \vdash \sigma: \T$.
  \item [(iii)] If $\Delta$ is a derivation of $\Gamma_1, t
    \match \tau, \Gamma_2 \vdash \mathcal{A}$, then there exists a sub-derivation
    $\Delta' \subseteq \Delta$ of $\Gamma_1 \vdash \tau : \T$.
  \end{itemize}
\end{lemma}  

\begin{lemma} (Weakening) \label{sec:weak}
    \begin{itemize}
    \item[(i)] If $\Gamma_1, \Gamma_2 \vdash \mathcal{A}$ and $\Gamma_1, \mathcal{C}, \Gamma_2
      \vdash ok$, then $\Gamma_1, \mathcal{C}, \Gamma_2 \vdash \mathcal{A}$.
    \item[(ii)] If $\Gamma_1 \vdash \mathcal{A}$ and $\Gamma_1, \Gamma_2 \vdash ok$,
      then $\Gamma_1, \Gamma_2 \vdash \mathcal{A}$.
  \end{itemize}
\end{lemma}  

\begin{lemma} (Well-formed object-types) \label{sec:type} 
  \begin{itemize}  
    \item[(i)] $\Gamma \vdash \proj{R}{m}:\T$ if and only
    if $\Gamma \vdash \pro{R}:\T$ and $\vm{} \subseteq \meth{R}$.
    
    \item[(ii)] $\Gamma \vdash t \ovatype \vm{}:\T$ if and only if  $\Gamma$
    contains $t \match \proj{R}{n}$, with $\vm{} \subseteq \meth{R}$.
\end{itemize}
\end{lemma}   

\begin{proposition} (Matching is well-formed) \label{sec:match-well}

\medskip
If $\Gamma \vdash \tau_1 \match \tau_2$, then
  $\Gamma \vdash \tau_1 : \T$ and $\Gamma \vdash \tau_2 : \T$.
\end{proposition}   

\begin{lemma} (Matching) \label{sec:matching} 
  \begin{itemize}  
  \item[(i)] $\Gamma \vdash \proj{R_1}{m} \match
    \tau_2$ if and only if $\Gamma \vdash \proj{R_1}{m}:\T$
    and $\Gamma \vdash \tau_2 : \T$ and $\tau_2 \equiv \proj{R_2}{n}$, 
    with $R_2 \subseteq R_1$ and $\vn{} \subseteq \vm{}$.

  \item[(ii)] $\Gamma \vdash \tau_1 \match t \ovatype \vn{}$ if and only if
    $\Gamma \vdash \tau_1 : \T$ and $\tau_1 \equiv t \ovatype \vm{}$, with
    $\vn{} \subseteq \vm{}$.

  \item[(iii)] $\Gamma \vdash t \ovatype \vm{} \match \proj{R_2}{n}$ if and only if $\Gamma$ contains $t \match \proj{R_1}{p}$, with $R_2 \subseteq R_1$ and $\vn{} \subseteq \vm{} \cup \vp{}$.

  \item[(iv)] (Reflexivity) If $\Gamma \vdash \rho : \T$ then $\Gamma
    \vdash \rho \match \rho$.

  \item[(v)] (Transitivity) If $\Gamma \vdash \tau_1 \match \rho$ and
    $\Gamma \vdash \rho \match \tau_2$, then $\Gamma \vdash \tau_1
    \match \tau_2$.

  \item[(vi)] (Uniqueness) If $\Gamma \vdash \tau_1 \match \probject{R_1, m\of\sigma_1}$ and $\Gamma \vdash \tau_1 \match \probject{R_2, m\of\sigma_2}$, then $\sigma_1 \equiv \sigma_2$.

  \item[(vii)] If $\Gamma \vdash \tau_1 \match \tau_2$ and 
      $\Gamma \vdash \tau_2 \tr \m  : \T$, then $\Gamma \vdash \tau_1 \tr \m
    \match \tau_2 \tr\m$.

  \item [(viii)] If $\Gamma \vdash \tau_1 \tr \m \match \proj{R}{n}$,
  then $\Gamma \vdash \tau_1 \match \proj{R}{n} {-} \m$.

\item [(ix)] If $\Gamma \vdash \rho \tr \m : \T$, then $\Gamma \vdash
  \rho \tr \m \match \rho$.
 \end{itemize}
\end{lemma}   

\begin{lemma} (Match Weakening) \label{sec:match-weak}
\begin{itemize}
\item[(i)] If $\Gamma_1, t \match \rho,\Gamma_2 \vdash \mathcal{A}$ and $\Gamma_1 \vdash \tau \match \rho$, with $\tau$ a {\tt pro}-type, then $\Gamma_1, t
\match \tau, \Gamma_2 \vdash \mathcal{A}$.
  
\item[(ii)] If $\Gamma \vdash \probject{R, \n\of\sigma} \ovatype \vm{} : \T$,
then $\Gamma, t \match \probject{R, \n\of\sigma} \ovatype \vm{} \vdash \sigma
: \T$.
\end{itemize}  
\end{lemma} 

\begin{proposition} (Substitution) \label{sec:subst}
  \begin{itemize}        
  \item [(i)] If $\Gamma_1,x\of\sigma,\Gamma_2 \vdash \mathcal{A}$ and $\Gamma_1
    \vdash \e :\sigma$, then $\Gamma_1,\Gamma_2 \vdash \mathcal{A}[\e/x]$.
    
  \item [(ii)] If $\Gamma_1, t \match \tau, \Gamma_2,\Gamma_3 \vdash \mathcal{A}$ and
    $\Gamma_1,t \match \tau, \Gamma_2 \vdash \rho \match \tau$, then
    $\Gamma_1, t \match \tau, \Gamma_2, \Gamma_3[\rho/t] \vdash
    \mathcal{A}[\rho/t]$.
  
  \item [(iii)] If $\Gamma_1, t \match \tau, \Gamma_2 \vdash \mathcal{A}$ and $\Gamma_1 \vdash \rho \match \tau$, then $\Gamma_1, \Gamma_2[\rho/t] \vdash
    \mathcal{A}[\rho/t]$.    
  \end{itemize}
\end{proposition}

\begin{proposition} (Types of expressions are well-formed) \label{sec:expr}
  
  \medskip
  If $\Gamma \vdash \e :\beta$, then $\Gamma
  \vdash \beta :\T $.
\end{proposition}

\noindent 
Finally, we can state the key Subject Reduction property for our type system.

\begin{theorem} (Subject Reduction, \ObjP) \label{sec:sr}
If $\Gamma \vdash \e :\beta$ and $\e \eval \e'$, then $\Gamma \vdash \e':\beta$.
\end{theorem}    

\noindent 
We proceed by deriving the Type Soundness theorem: it
guarantees, among other properties, that every closed and well-typed expression
will not produce the \emph{message-not-found} runtime error.  This
error arises whenever we search for a method $\m$ into an expression that
does not reduce to an object which has the method $\m$ in its
interface.

\begin{definition} 
We define the set of $\wrong$ terms as follows:
\[
\begin{array}{lcl}
\wrong & ::= & Sel(\l \, \rangle, \m, \e)  \mid 
               Sel((\lambda x.\e), \m, e') \mid 
               Sel(c, \m, \e)
\end{array}
\]
\end{definition}
By a direct inspection of the typing rules for terms, one can
immediately see that $\wrong$ cannot be typed. Hence, the Type 
Soundness follows as a corollary of the Subject Reduction theorem.

\begin{corollary} (Type Soundness) \label{sec:ts}
  If $\varepsilon \vdash \e : \beta$, then
  $\e \not \!\!\!\! \deval C[\wrong]$, where $C[\;]$ is a generic context
  in \ObjP, i.e.  a term with an ``hole'' inside it.
\end{corollary}


\subsection{Soundness of the Type System with Subsumption}\label{sec:soundness2}

The proof of the Type Soundness concerning the type assignment system with subsumption \ObjPS\ is quite similar to the corresponding proof for the
plain type system.
In particular, all the preliminary lemmas and their corresponding proofs remain almost the same; only the proof of the crucial Theorem \ref{sec:sr} needs to be modified significantly.
Therefore, we do not document the whole proofs of the preliminary lemmas, but we just remark the points where new arguments are needed.

In fact, Lemmas \ref{sec:sub} (Sub-derivation), \ref{sec:weak} (Weakening),
\ref{sec:match-well} (Matching is well-formed), \ref{sec:subst}
(Substitution), \ref{sec:expr} (Types of expressions are well-formed) are
valid also for the type assignment with subsumption.
Conversely, we need to rephrase Lemmas \ref{sec:type} (Well-formed
object-types), \ref{sec:matching} (Matching), \ref{sec:match-weak} (Match Weakening), as follows.

In Lemma (Well-formed object-types) \ref{sec:type}, the point (ii) needs to be
rewritten as:
\begin{itemize} 
\item[(ii)] $\Gamma \vdash t \ovatype \vm{}:\T$ if and only
 if $\Gamma$ contains either $t \match \proj{R}{n}$ or
 $t \match \objj{R}{n}$, with $\vm{} \subseteq \meth{R}$.
\end{itemize}
In Lemma (Matching) \ref{sec:matching}, the point (vi) needs to be rewritten as:
\begin{itemize}
\item[(vi)] (Uniqueness) if $\Gamma \vdash \tau_1 \match \object{R_1, m\of\sigma_1}$ and $\Gamma \vdash \tau_1 \match \object{R_2, m\of\sigma_2}$, then $\sigma_1 \equiv \sigma_2$.
\end{itemize}
Moreover, in the same lemma the following points need to be added:
  \begin{itemize}
  \item[(i')] $\Gamma \vdash \objj{R_1}{m} \match
    \tau_2$ if and only if $\Gamma \vdash \proj{R_1}{m}:\T$ and
    $\Gamma \vdash \tau_2 : \T$ and
    $\tau_2 \equiv \objj{R_2}{n}$, 
    with $R_2 \subseteq R_1$ and $\vn{} \subseteq \vm{}$.
      
  \item[(iii')] $\Gamma \vdash t \ovatype \vm{} \match \objj{R_2}{n}$
  if and only if $\Gamma$ contains either $t \match \objj{R_1}{p}$ or
  $t \match \proj{R_1}{p}$, with
  $R_2 \subseteq R_1$ and $\vn{} \subseteq \vm{} \cup \vp{}$.
 
  \item [(viii')] If $\Gamma \vdash \tau_1 \tr \m \match \objj{R}{n}$, then
  $\Gamma \vdash \tau_1 \match \objj{R}{n} {-} \m$.
\end{itemize}
In Lemma \ref{sec:match-weak} (Match Weakening), the point (ii) needs to be rewritten as:
\begin{itemize}
 \item[(ii)] If $\Gamma \vdash \probject{R, \n\of\sigma} \ovatype \vm{} : \T$ or 
 $\Gamma \vdash \object{R, \n\of\sigma} \ovatype \vm{} : \T$ can be derived,
then $\Gamma, t \match \object{R, \n\of\sigma} \ovatype \vm{} \vdash \sigma
: \T$.
\end{itemize}  
A new lemma, stating some elementary properties of types with 
covariant variables and rigid types is necessary.
\begin{lemma}  (Covariant variables and rigid types) \label{rigid}
\begin{itemize}
\item[(i)]
If $t$ is covariant in $\sigma$ and $\Gamma \vdash \sigma : \TRIG$ and
$\Gamma \vdash \tau_1 \match \tau_2$, then
$\Gamma \vdash  \sigma[\tau_1/t] \match \sigma[\tau_2/t]$.

\item[(ii)]
If $\Gamma \vdash \sigma_1 : \TRIG$ and $\Gamma \vdash \sigma_2 : \TRIG$, then  $\Gamma \vdash \sigma_1[\sigma_2/t] : \TRIG$.
\end{itemize}
\end{lemma}
Finally, the Subject Reduction for the type assignment system with subsumption has the usual formulation, but needs a more complex proof (reported in Appendix \ref{soundness2}).

\begin{theorem} (Subject Reduction, \ObjPS) \label{sec:sr-full}
If $\Gamma \vdash \e :\beta$ and $\e \eval \e'$, then $\Gamma \vdash \e':\beta$.
\end{theorem}
  


\section{Object reclassification}\label{sec:reclass}

The natural counterpart of self-extension in class-based languages is known as ``(dynamic) object reclassification''.
This operation allows for the possibility of changing at runtime the class membership of an object while retaining its identity.
%
%
%
One major contribution to the development of reclassification features has produced the Java-like Fickle language, in its incremental versions \cite{fickle1:fool01,fickle2:toplas02,fickle3:ictcs03}.

In this section, we show how the self-inflicted extension primitive provided by our calculus may be used to mimic the mechanisms implemented in Fickle.
We proceed, suggestively, by working out a case study: first we write an example in Fickle which illustrates the essential ingredients of the reclassification, then we devise and discuss the possibilities of its encoding in \ObjP.
 
\subsection{Reclassification in Fickle}\label{subsec:fickle}

Fickle is an imperative, class-based, strongly-typed language, where classes are types and subclasses are subtypes.
It is statically typed, via a type and effect system which turns out to be sound w.r.t. 
the operational semantics.
Reclassification is achieved by dynamically changing the class membership of objects; correspondingly, the type system guarantees that objects will never access non-existing class components.

To develop the example in this section, we will refer to the second version of the language (known as Fickle$_{II}$ \cite{fickle2:toplas02}).



In the Fickle scenario, an abstract class {\tt C} has two non-overlapping 
concrete subclasses {\tt A} and {\tt B}, where the three classes must be of two 
different kinds: {\tt C} is a root class, whereas {\tt A} and {\tt B} are state ones.
In fact, one finds in root classes, such as {\tt C}, the declaration of the (private) 
attributes (a.k.a. fields) and the (public) methods which are common to its state subclasses.
On the other hand, state classes, such as {\tt A} and {\tt B}, are intended to serve 
as targets of reclassifications, hence their declaration contains the extra attributes 
and methods that exclusively belong to each of them.


The reclassification mechanism allows one object in a state class, say {\tt A}, to 
become an object of the state class {\tt B} (or, viceversa, moving from {\tt B} to {\tt A}) through the execution of a reclassification expression.
The semantics of this operation, which may appear in the body of methods, is that the attributes of the object belonging to the source class are removed, those common to the two classes (which are in {\tt C}) are retained, and the ones belonging to the target class are added to the object itself, without changing its identity.
The same happens to the methods component, with the difference that the abstract 
methods declared in {\tt C} (therefore common to {\tt A} and {\tt B}) may have different bodies in the two subclasses: when this is the case, reclassifying an object means replacing the bodies of the involved methods, too.

In the example of Figure \ref{fickle-person}, written in Fickle syntax, we first introduce the class \texttt{Person}, with an attribute to name a person and an abstract method to employ him/her.
Then we add two subclasses, to model students and workers, with the following intended meaning.
The \texttt{Student} class extends \texttt{Person} via a registration number (\texttt{id} attribute) and by instantiating the \texttt{employment} method.
The \texttt{Worker} class extends \texttt{Person} via a remuneration information (\texttt{salary} attribute), a different \texttt{employment} method, and the extra \texttt{registration} method to register as a student.
We remark that, in our example, students and workers are mutually exclusive.

\begin{figure}[t]
\begin{alltt}
abstract root class Person extends Object \{
 string name;
 abstract void employment(int n) \{Person\};
\}

state class Student extends Person \{
 int id;
 Student(string s, int m) \{ \} \{name:=s; id:=m\};
 void employment(int n) \{Person\} \{this=>Worker; salary:=n\};
\}

state class Worker extends Person \{
 int salary;
 Worker(string s, int n) \{ \} \{name:=s; salary:=n\};
 void employment(int n) \{ \} \{salary:=salary+n\};
 void registration(int m) \{Person\} \{this=>Student; id:=m\};
 \}
\end{alltt}
\vspace{-3mm}
\caption{Person-Student-Worker example}
\label{fickle-person}
\end{figure}

The root class \texttt{Person} defines the attributes and methods common to its 
state subclasses \texttt{Student} and \texttt{Worker} (notice that, being its 
\texttt{employment} method abstract, the root class itself must be abstract, 
therefore not supplying any constructor).


The classes \texttt{Student} and \texttt{Worker}, being subclasses of a root one (i.e. \texttt{Person}), must be state classes, which means that may be used as targets of reclassifications.
Annotations, like \texttt{\{\ \}} and \texttt{\{Person\}}, placed before the bodies of the methods, are named effects and are intended to list the root classes of the objects that may be reclassified by invoking those methods: in particular, the empty effect \texttt{\{\ \}} cannot cause any reclassification and the non-empty effect \texttt{\{Person\}} allows to reclassify objects of its subclasses.
Let us now consider the following program fragment:
\begin{alltt}
1. Person p,q;
2. p := new Student("Alice",45);
3. q := new Worker("Bob",27K);
\end{alltt}
After these lines, the variables \texttt{p} and \texttt{q} are bound to a \texttt{Student} and a \texttt{Worker} objects, respectively.
To illustrate the key points of the reclassification mechanism, we make {\tt Bob} become a {\tt Student}, and {\tt Alice} first become a {\tt Worker} and then get a second job:
\begin{alltt}
4. q.registration(57); 
5. p.employment(30K);
6. p.employment(14K);
\end{alltt}
Line $4$, by sending the \texttt{registration} message to the object \texttt{q}, 
causes the execution of the reclassification expression \texttt{this=>Student}: before its execution, the receiver \texttt{q} is an object of the \texttt{Worker} class, therefore it contains the \texttt{salary} attribute; after it, \texttt{q} is reclassified into the \texttt{Student} class, hence \texttt{salary} is removed, \texttt{name} is not affected, and the \texttt{id} attribute is added and instantiated with the actual parameter.

Coming to the second object \texttt{p}, belonging to \texttt{Student} and representing {\tt Alice}, line $5$ carries out exactly the opposite operation w.r.t. line $4$, by reclassifying \texttt{p} into the \texttt{Worker} class via the expression \texttt{this=>Worker}, with the result that \texttt{id} is no longer available, \texttt{name} preserves its value, and \texttt{salary} is added and instantiated.

The following line $6$, therefore, selects the \texttt{employment} method from 
\texttt{Worker}, not from \texttt{Student} as before, because the object \texttt{p} 
has been reclassified in the meantime.
This latter invocation of \texttt{employment} has the effect of augmenting {\tt Alice}'s income by the actual parameter value, thus allowing us to model a sort of multi-worker.

\subsection{Desiderata}
\label{subsec:goals}

In this section, we devise the ``ideal'' behaviour of \ObjP\ w.r.t. the reclassification goal, without guaranteeing that the terms we introduce can be typed. 

It is apparent that the main tool provided by our calculus to mimic Fickle's reclassification mechanism is the self-extension primitive; precisely, we need a \emph{reversible} extension functionality, to be used first to extend an object with new methods and later to remove from the resulting object some of its methods.
Hence, an immediate solution would rely on a massive use of the self-extension primitive, as follows:
$$
\begin{array}{lcrcl}
alice & \triangleq &
\l name 
& = & \textrm{``Alice''},
\\
& &
reg 
& = & 
\lambda s.\lambda m.\l \l s \ova id = m \r
\\
& &
& & \phantom{\lambda s.\lambda m.\l \l s}
\ova emp = \lambda n.s \call emp(n) \r,
\\
& & 
emp 
& = &
\lambda s.\lambda m. \l\l\l s \ova sal = m \r 
\\
& &
& &
\phantom{\lambda s\lambda m. \l\l\l s\;} \ova reg = \lambda n. s\call reg(n) \r
\\
& &
& &
\phantom{\lambda s\lambda m. \l\l\l s\;} \ova emp = \lambda s'.\lambda p. \l s'\ova sal = (s' \call sal) + p \r\r\r
\end{array}
$$
To model the example of Figure \ref{fickle-person} in \ObjP, we have defined the $alice$ object prototype for representing Alice as a person.
Now, it can be extended to either a student or a worker via the $reg$ (i.e. registration) or $emp$ (i.e. employment) methods, which are intended to play the role of the \texttt{Student} and \texttt{Worker} constructors of Section \ref{subsec:fickle}, respectively.
We illustrate the behaviour of the former;
in fact, $alice$ becomes a student through the $reg$ method, which adds $id$ to the receiver and overrides the $emp$ method. Therefore, $alice \call reg(45)$ reduces to the following object:
$$
\begin{array}{lcrcl}
alice_S & \triangleq &
\l name,reg,emp 
& =& 
\mbox{as in $alice$},
\\ 
& &
id 
& = & 
45,
\\
& &
emp 
& = & 
\lambda m.alice \call emp(m)
\r
\end{array}
$$
In this way, the prototype $alice$ is stored in the body of the novel $emp$ method in the perspective of a reclassification:
no matter if a cascade of $reg$ is invoked and $emp$ methods are stacked, because eventually the present version of $emp$ is executed\footnote{An alternative solution would be that $reg$ in $alice$ overrides itself as $reg = \lambda s'.\lambda p. \l s' \ova id = p \r$; in such an equivalent case only $id$ methods would be stacked, rather than $\langle id, emp \rangle$ pairs.}.

Then, $alice_S$ can be reclassified into a worker via the invocation of such an $emp$, which sends to the original $alice$ its former version (i.e. $alice$'s third method). In fact, $alice_S \call emp(30K)$ reduces to:
$$
\begin{array}{lcrcl}
alice_W & \triangleq &
\l name,reg,emp  
& = & 
\mbox{as in $alice$},
\\
& &
sal 
& = & 
30K,
\\
& &
reg 
& = & 
\lambda m. alice \call reg(m),
\\ 
& &
emp 
& = & 
\lambda s.\lambda n.  \l s \ova sal = (s \call sal) + n \r 
\r 
\end{array}
$$
As the reader can see, the effect of this message is that the methods characterizing a student are removed (by coming back to $alice$) and those needed by a worker, in turn, extend $alice$; notice that the novel version of $emp$ models the multi-worker.

To finalize the modeling of Section's \ref{subsec:fickle} example in our calculus, $alice_W$'s income may be increased by means of a call to such a version of $emp$, which has overridden $alice$'s third method; i.e. $alice_W \call emp(14K)$ reduces to:
$$
\begin{array}{lcrcl}
alice_{W_2} & \triangleq &
\l name,reg,emp,sal,reg,emp 
& = & 
\mbox{as in $alice_W$},
\\
& &
sal 
& = & 
(alice_W \call sal) + 14K 
\r
\end{array}
$$
\paragraph{About typability.} The encoding devised in this section may be seen as a reasonable solution to emulate object reclassification in \ObjP; however, the actual free use of the self-extension primitive does not allow us to type the terms introduced.

The point is that the self-variable, representing the receiver object, cannot be used in the body of a method added by self-extension to remove methods, in the attempt to restore the receiver before its extension (it is the case of $emp$'s body, added by the second method $reg$ and, symmetrically, $reg$'s body, added by $emp$).

We can discuss the issue via the minimal (hence simpler than $alice$) object:
$$
andback \ \triangleq \ \l extend = \lambda s. \l s \ova delete = \lambda s'. s \r \r 
$$
The difficulty to type $andback$ concerns the type returned by the $delete$ method:
$$
andback \ : \ 
\probject{extend \of  t \ovatype delete,\ delete \of \mathbf{?}} \ovatype extend
$$

We first observe that the type variable $t$ would not be a suitable candidate for $delete$, because $t$, within the scope of the above $pro$ binder, is intended to represent the receiver, i.e. in the $delete$ case at hand, the object \emph{already} extended and therefore containing the $delete$ method.

A second attempt would be typing $andback$ itself with the type
returned by $delete$:
$$
andback \ : \ \tau \ \triangleq\
\probject{extend \of t \ovatype delete,\ delete \of \tau} \ovatype extend 
$$
That is, the candidate type $\tau$ should satisfy a recursion equation.
However, \ObjP's recursion mechanisms is not powerful enough to express such a type, hence we are devoting the remaining part of Section \ref{sec:reclass} to design alternative and typable encodings.

\subsection{The runtime solution}\label{subsec:extension}

A first possibility to circumvent the tipability problem arised in Section \ref{subsec:goals} is very plain: at first we extend an object with new methods, and from then we keep just overriding the resulting object, without removing methods from it.
That is, the first use of the self-extension leads to object extension, whereas all the following ones to object override.
We may then model Figure \ref{fickle-person}'s example via the following prototype:
\[
\begin{array}{lcrcl}
alice' & \triangleq &
\l  name 
& = & \textrm{``Alice''},
\\
& &
reg 
& = & 
\lambda s.\lambda m. \l \l s \ova id = m \ra \ova sal = 0 \ra,
\\
& & 
emp 
& = & \lambda s.\lambda m. \l  \l  \l  s \ova id = 0 \ra \ova sal = m \ra
\\
& &
&  & \phantom{\lambda s.\lambda m. \l  \l  \l  s}
\ova emp = \lambda s'.\lambda n. \l \l s' \ova id =0 \ra
\\
& &
& & \phantom{\lambda s.\lambda m. \l  \l  \l  s \ova emp = \lambda s'.\lambda n. \l \l s'}
\ova sal = (s' \call sal) + n \ra \ra \ra
\end{array}
\]
As the reader can inspect, in this alternative Alice's encoding the variables representing the host object ($s$ and $s'$) are never used in a method body to represent the receiver without the method being defined.
This crucial fact holds also for the rightmost $sal$, where $s'$ refers to an object where that method is already available; such a property can be checked syntactically, hence $alice'$ may be given the following type:
\begin{equation}\label{runtime-type}
\begin{array}{lcrcl}
alice' & : &
\probject{
name
& : & String,
\\
& &
reg 
& : & \mathbb{N} \to t \ovatype id \ovatype sal,
\\
& &
emp 
& : & \mathbb{N} \to t \ovatype id \ovatype sal,
\\
& &
id 
& : & \mathbb{N},
\\
& &
sal 
& : & \mathbb{N}
} \tr name,reg,emp 
\end{array}
\end{equation}

The price to pay for typability is that the objects playing the roles of students and workers will contain more methods than needed (all the methods involved), because no method can be removed. In the present example, when $alice$' registers as a student, $id$ and $sal$ are added permanently to the interface, i.e. $alice' \call reg(45)$ reduces to:
$$
\begin{array}{lcrcl}
alice'_S & \triangleq &
\l name,reg,emp 
& =& 
\mbox{as in $alice'$},
\\ 
& &
id 
& = & 
45,
\\
& &
sal
& = & 
0
\r
\end{array}
$$
%
Therefore, the type system will not detect type errors related to uncorrect method calls.
In fact, $alice'_S$ is intended to represent a student, but in practice we will have to distinguish between students and workers via the runtime answers to the $id$ and $sal$ (representing students' and workers' attributes, respectively) method invocations: non-zero values (such as $45$, returned by $id$) are informative of genuine attributes, while zero values (returned by $sal$) tell us that the corresponding attribute is not significant.
This solution is reminiscent of an approach to reclassification via wide classes, requiring runtime tests to diagnose the presence of fields \cite{Serrano-ecoop99}.

We proceed by reclassifying $alice'_S$ into a worker; $alice'_S \call emp(30K)$ reduces to\footnote{Notice that, to ease readability, we will omit from now on the overriden methods, if the latter have definitively become garbage (in the case: the inner versions of $emp$, $id$, $sal$).}:
\[
\begin{array}{lcrcl}
alice'_W & \triangleq &
\l name,reg  & = & \mathrm{as\ in}\ alice',
\\ 
& & 
id & = & 0,
\\
& &
sal & = & 30K,
\\
& & 
emp & = &
\lambda s.\lambda m. \l \l s  \ova id = 0 \ra
\\
& &
& & \phantom{\lambda s.\lambda m. \l \l s}
\ova sal = (s \call sal) + m \ra \ra
\end{array}
\]
The consequence of this call to (the original) $emp$ is that $id$ and $sal$ swap their role, thus making effective the reclassification, and a new version of $emp$ is embedded in the interface.
Notice that such a novel $emp$ (incrementing the salary $sal$) works correctly not only with the usual multi-worker operation $alice'_W \call emp(14K)$, reducing to:
\[
\begin{array}{lcrcl}
alice'_{W_2} & \triangleq &
\l name,reg,emp & = & \mathrm{as\ in}\ alice'_W,
\\
& &
id & = & 0,
\\
& &
sal & = & (alice'_W \call sal) + 14K \ra
\\
\end{array}
\]
but also in the case of a further reclassification of $alice'_W$ into a student, because setting ex-novo a salary is equivalent to adding it to the zero value stored by $reg$.

Finally, a couple of remarks about the relationship of the two $emp$ versions with the type (\ref{runtime-type}).
First, the fact that the overridden $emp$ (i.e. the one belonging to $alice'$) extends the receiver via $id$ and $sal$ but overrides itself is clearly expressed by its type $\mathbb{N} \to t \ovatype id \ovatype sal$.
Second, the redundant $id$ version contained by the overriding $emp$ (i.e. the one that appears in $alice'_W$) is hence necessary to respect such a type.

\subsection{Creating new objects}\label{subsec:newobjects}
A second way to achieve the possibility to remove methods from an object is by creating new objects.
To illustrate such an approach, we pick out the following object:
\[
\begin{array}{lcl}
andback' & \triangleq &
\l extend = \lambda s. \l extend = \lambda s'. s', delete = \lambda s'. s \ra \ra
\end{array}
\]
which models the same behavior of the minimal $andback$, introduced in Section \ref{subsec:goals} to enlighten the typability problem that we want to encompass.
In the present case, the method $delete$ is allowed by the type system to return its prototype object (represented by the variable $s$), because such a method belongs to a completely \emph{new object}, not to an object which has extended its prototype (as it was in Section \ref{subsec:goals}):
%
%
$$
andback' \ : \ \tau' \ \triangleq\
\probject{extend \of  \probjecu{extend \of  t',delete \of  t} \ovatype extend,delete} \ovatype extend
$$
The reader may observe how the type $\tau'$ reflects the explanation given above: a new object is generated via the $extend$ method and represented by $t'$; within such an object, the $delete$ method refers to the prototype object, represented by $t$.

We apply the idea to our working example; combining the self-extension primitive with the generation of new objects leads to a third Alice's representation:
%
\[
\begin{array}{lcrcl}
alice'' & \triangleq &
\l  name 
& = & \textrm{``Alice''},
\\
& &
reg 
& = & 
\lambda s.\lambda m.\l name = s \call name,
\\
& &
& &	\phantom{\lambda s.\lambda m.\l}
id = m,
\\
& &	
& &	\phantom{\lambda s.\lambda m.\l}
emp = \lambda n. \l  s \ova sal = 0  \ra \call emp(n)  \ra,
\\	
& &
emp 
& = & 
\lambda s.\lambda m. \l  \l  s \ova sal = m \ra\\
& &
& & \phantom{ \lambda s.\lambda m. \l  \l s}
      \ova emp = \lambda s'.\lambda n. \l  s' \ova sal = (s' \call sal) + n \ra  \ra  \ra
\end{array}
\]
The novelty of the present solution amounts to the fact that the $reg$ method creates a new object from scratch, equipped with three methods: the first one copies the $name$ value from its prototype, the second method sets the $id$ attribute, and, the key point, the $emp$ method is allowed to refer back to the prototype object to prepare for a potential worker reclassification.
As argued above, this latter method is typable, conversely to its version in $alice$ (Section \ref{subsec:goals}), because it is not added by self-extension, but belongs to a different object, created ex-novo.
In the end, the $alice''$ prototype object can type-checked against the following type\footnote{Typing the third method $emp$ is not problematic, being simpler than in previous Section \ref{subsec:extension}.}:
%
\[
\begin{array}{lcrcl}
alice'' \ : \ \rho & \triangleq &
\probject{
name & : &
String,
\\
& &
reg & : &
\mathbb{N} \to \probjecu{name : String,
\\
& & & &
\phantom{\mathbb{N} \to {\tt pro}\ssk t'.\l}
id : \mathbb{N},
\\
& & & &
\phantom{\mathbb{N} \to {\tt pro}\ssk t'.\l}
emp : \mathbb{N} \to t \ovatype sal
} \ovatype name,id,emp,
\\  
& &
emp & : & 
\mathbb{N} \to t \ovatype sal,
\\
& &
sal & : & 
\mathbb{N}} \ovatype  name,reg,emp\\
\end{array}
\]
where it is apparent that both the $emp$ versions add $sal$ to $alice''$'s interface.
Then, the outcome of Alice's registration, $alice'' \call reg(45)$, is the following:
\[
\begin{array}{lcrcl}
alice''_S & \triangleq & \l
name & =& alice'' \call name,
\\
& & 
id & = & 45,
\\
& & 
emp & = & \lambda m. \l  alice'' \ova sal = 0 \ra \call emp(m) \ra
\\
\\
alice''_S & : &
\probjecu{
name & : & String,
\\
& &
id & : & \mathbb{N} ,
\\
& &
emp & : & \mathbb{N} \to \rho \ovatype sal} \ovatype name,id,emp
\end{array}
\]
One can see in this latter type that, coherently, the $emp$ method adds $sal$ to the prototype $alice''$.
We observe also that, in $emp$'s body, a ``local'' version of $sal$ is added on the fly to the receiver ($alice''$, in the case) before the call to the outer $emp$.
This is necessary to guarantee the correctness of the protocol in the event of a call to $alice''$'s $emp$ before than $reg$ (an example that we do not detail here):
$emp$ overrides itself, thus losing from then the possibility to set the salary from scratch (see the $alice''$ term), which must be hence incremented starting from zero.

The chance to send $emp$ to the prototype $alice''$, via the $alice''_S \call emp(30K)$ call, is crucial for the reclassification, giving in fact the following outcome:
\[
\begin{array}{lcrcl}
alice''_W & \triangleq & \l 
name,reg & = & \textrm{as in $alice''$},
\\
& & 
sal & = & 30K,
\\
& & 
emp & = & \lambda s.\lambda m. \l  s \ova sal =(s \call sal) + m \ra  \ra
\\
\\
alice''_W & : &
\probject{
name & : &
String,
\\
& & reg & : & 
\mathbb{N} \to \probjecu{name : String,
\\
& & & &
\phantom{\mathbb{N} \to \proa \l} id : \mathbb{N},
\\
& & & &
\phantom{\mathbb{N} \to \proa \l} emp : \mathbb{N} \to t } \ovatype name,id,emp
\\
& &
sal  & : & 
\mathbb{N},
\\
& &
emp & : & 
\mathbb{N} \to t} \ovatype 
name,reg,sal,emp\\
\end{array}
\]
where the presence of the salary in the new interface is reflected by both $emp$'s types.

We end by adding the usual second job to Alice, through the $alice''_W \call emp(14K)$ call, which reduces to the following object, whose type is the same of $alice''_W$:
\[
\begin{array}{lcrcl}
alice''_{W_2} & \triangleq & \l 
name,reg,emp & = & \textrm{as in $alice''_W$},
 \\
& & 
sal & = & (alice''_W \call sal) + 14K  \ra
\\
\end{array}
\]
\paragraph{Discussion.} It is apparent that the opposite reclassification direction (Alice first becoming a worker and then a student) would produce terms behaviourally equivalent to $alice''_W$ and $alice''_S$, even though not syntactically identical.

We remark also that in fact a couple of choices is already feasible, if one decides to combine self-extensions and new objects: in principle, there is no reason to prefer the encoding that we have illustrated to the symmetrical one (simpler, in the case), where students are modeled via self-extensions and workers through new objects.

To conclude, the reader might wonder about the asymmetry of the solution developed in this section, as students are managed via new objects and workers through self-extensions.
Actually, in Section \ref{subsec:goals} we have shown that modeling the reclassification by means of the sole self-extension mechanism leads to non-typable terms.
On the opposite side, it is always possible to encode the reclassification via only new objects (to manage also workers), without the need of the self-extension:
%
\[
\begin{array}{lcrcl}
alice''' & \triangleq & \l name 
& = &
\textrm{``Alice''},
\\
& &
reg 
& = & 
\lambda s.\lambda m. \l name = s \call name,\\
&&&&
\phantom{\lambda s.\lambda m. \l }
		id = m,\\
&&&&\phantom{\lambda s.\lambda m. \l }
		emp = \lambda n. s \call emp(n)  \ra,
\\
& &
emp 
& = & 
\lambda s.\lambda m. \l
	name = s \call name,\\
&&&& \phantom{\lambda s.\lambda m. \l }
	sal = m,\\
&&&& \phantom{\lambda s.\lambda m. \l }
	emp = \lambda s'.\lambda n. \l  s' \ova sal =(s' \call sal) + n \ra,\\
&&&& \phantom{\lambda s.\lambda m. \l }
	reg = \lambda p. s \call reg(p)  \ra  \ra\\
\end{array}
\]
Summarizing, in this section we have tried to push the self-extension, which is the technical novelty of this paper, to its limit (i.e. typability).
We believe that such an effort is interesting per se;
moreover, the ``mixed'' solution which arises from our investigation leads to a more compact encoding, giving the benefit of code reuse.




\section{Related work}\label{related}

Several efforts have been carried out in recent years with an aim similar to that of our work, namely for the sake of providing static type systems 
for object-oriented languages that change at runtime the behaviour of objects.
In this section, first we discuss the approaches in the literature by 
considering separately the two main categories of prototype-based 
and class-based languages, afterwards we survey the relationship 
between object extension and object subsumption.

\subsection{In prototype-based languages}\label{related-prototypes}
A few works consider the problem of defining static type disciplines for JavaScript, a prototype-based, dynamically typed language where objects can be modified at runtime and errors caused by calls to undefined methods may occur.


Zhao in \cite{Zhao12} presents a static type inference 
algorithm for a fragment of JavaScript and suggests two 
type disciplines for preventing undefined method calls.
Similarly to the \ObjP\ calculus, JavaScript provides self-inflicted 
extension; to deal with this feature, some ideas shared with our 
approach are adopted, namely
$i)$ the distinction between $pro$-types and $obj$-types,
$ii)$ the distinction between ``available'' and ``reserved'' methods, and
$iii)$ the mechanisms to mark the migration of a method from reserved to available.
On the other hand, the main differences or extra features w.r.t. our work are the following:
$a)$ JavaScript allows strong update, i.e. overriding a method with a different type, and the type system accommodates, in a limited way, this functionality; 
$b)$ the types are defined by means of a set of subtyping constraints;
$c)$ the syntax is completely different. 

Chugh and co-workers propose in \cite{ChughHJ12} a static type system for 
quite a rich subset of JavaScript.
The considered features are imperative updates (i.e. updates that change 
the set of methods of an object by adding and also subtracting methods) and 
arrays, which in JavaScript can be homogeneous (when all the elements 
have the same type) but also heterogeneous, like tuples.
As the syntax makes no distinction between these two kinds of arrays, 
to form the correct type can be challenging.
In order to deal with subtyping and inheritance, the authors further elaborate 
our idea of splitting the list of methods into reserved and available parts.

Vouillon presents in \cite{Vouillon01} a prototype-based calculus containing 
the ``object-view'' mechanism, which permits to change the interface between an object and the environment, thus allowing an object to hide part of its methods 
in some context. As in our work, the author defines a distinction between 
$pro$-types and $obj$-types.

\subsection{In class-based languages}

The typical setting where class-based languages are investigated is a Java-like environment.
In the previous Section \ref{sec:reclass} we have considered object reclassification, a feature introduced in the class-based paradigm, and we have experimented with modeling in \ObjP\ the reclassification mechanism implemented in Fickle \cite{fickle2:toplas02}.
%
We complete now the survey of the involved related work by presenting other 
contributions that fall in the same class-based category.

Cohen and Gil's work \cite{CohenG09}, about the introduction of \emph{object evolution} into statically typed languages, is much related to reclassification, because evolution is a restriction of reclassification, by which objects may only gain, but never lose their capabilities (hence it may be promptly mimicked in \ObjP).
An evolution operation (which may be of three non-mutually 
exclusive variants, based respectively on inheritance, mixins, and shakeins) takes at runtime an instance of one class and replaces it with an instance of a selected subclass.
The monotonicity property granted by such a kind of dynamic change makes easier 
to maintain static type-safety than in general reclassification. 
In the end, the authors experiment with an implementation of evolution in Java, 
based on the idea of using a forward pointer to a new memory address to support the objects which have evolved, starting from the original non-evolved object.
%

Monpratarnchai and Tamai \cite{EpsilonJ} introduce an extension of Java named EpsilonJ, featuring role modeling (that is, a set of roles to represent collaboration carried out in that context, e.g. between an employer and its employees) and object adaptation (that is, a dynamic change of role, to partecipate in a context by assuming one of its roles).
Dynamically acquired methods obtained by assuming roles have to be invoked by means of down-casting, which is a type unsafe operation.
%
Later, Kamina and Tamai \cite{KaminaT10} introduce an extension of Java named NextEJ, to combine the object-based adaptation mechanisms of 
EpsilonJ and the object-role binding provided by context-oriented languages.
In fact, the authors model in  NextEJ the \emph{context activation scope}, 
adopted from the latter languages, and prove that such a mechanism is type sound 
by using a small calculus which formalizes the core features of NextEJ. 

Ressia and co-workers \cite{RessiaGNPR14} introduce a new form of inheritance called \emph{talents}. 
A talent is an object belonging to a standard class, named \texttt{Talent}, which can 
be acquired (via a suitable \texttt{acquire} primitive) by any object, which is then adapted. 
The crucial operational characteristics of talents are that they are scoped dynamically 
and that their composition order is irrelevant.
However, when two talents with different implementations of the same method are 
composed a conflict arises, which has to be resolved either through aliasing (the name of the method in a talent is changed) or via exclusion (the method is removed from a talent before composition). 


\subsection{Object extension vs. subsumption}

Several calculi proposed in the literature combine object extension with 
object subsumption.
Beside of the peculiar technicalities of those proposals, they all share the principle 
of avoiding (type incompatible) object extensions in presence of a (limited) form of 
object subsumption. 

Riecke and Stone in \cite{RS-98} present a calculus where it is possible to
first subsume (forget) an object component, and then re-add it again with a type which may be incompatible with the forgotten one.
In order to guarantee the soundness of the type system, method dictionaries are used inside objects with the goal of linking correctly method names and method bodies. 

Ghelli in \cite{Ghelli02} pursues the same freedom (of forgetting a method and 
adding it again with a different incompatible type) by introducing a context-dependent behaviour of objects called \emph{object role}.
Ghelli introduces a role calculus, which is a minimal extension of Abadi-Cardelli's $\varsigma$-calculus, where an object is allowed to change dynamically identity while keeping static type checking. 
Vouillon's ``view'' mechanism \cite{Vouillon01}, see Section \ref{related-prototypes}, can also be interpreted as a a kind of role.

Approaches to subsumption similar to the one presented in this work
can be found in \cite{FM-95,Liquori-Ecoop-97,BBDL-97,Remy-98}.
In \cite{Liquori-Ecoop-97}, an extension of Abadi-Cardelli's Object Calculus is
presented; roughly speaking, we can say that $pro$-types and $obj$-types in
the present article correspond to ``diamond-types'' and ``saturated-types'' in that work. 
Similar ideas can be found in \cite{Remy-98}, although the type system there presented permits also a form of self-inflicted extension. However, in that type system, a method $m$ performing a self-inflicted extension needs to return a rigid
object whose type is fixed in the declaration of the body of $m$.
As a consequence, the following expressions would not be typable in that system: 
$$
\begin{array}{l}
\l \l \p \ova new_m = \ldots \rangle\call \addsetc \rangle \call new_m \\ [2mm]
\l \l \p \call \addsetc \l \ova new_m = \ldots \rangle
\end{array}
$$
Another type system for the \Obj\ calculus is presented in \cite{BBDL-97};
such a type system uses a refined notion of
subtyping that allows to type also {\em binary methods}.


\backmatter



\newcommand{\etalchar}[1]{$^{#1}$}


\begin{thebibliography}{DDDG02}

\bibitem[AC96]{AC-book}
M.~Abadi and L.~Cardelli.
\newblock {\em A {T}heory of {O}bjects}.
\newblock 1996.

\bibitem[BB99]{BB-97}
Viviana Bono and Michele Bugliesi.
\newblock Matching for the lambda calculus of objects.
\newblock {\em Theoretical Computer Science}, 212(1-2):101--140, 1999.

\bibitem[BBDL97]{BBDL-97}
V.~Bono, M.~Bugliesi, M.~{Dezani-Ciancaglini}, and L.~Liquori.
\newblock Subtyping {C}onstraint for {I}ncomplete {O}bjects.
\newblock In {\em Proc. of TAPSOFT/CAAP}, volume 1214, pages 465--477, 1997.

\bibitem[BBL96]{BBL-96}
V.~Bono, M.~Bugliesi, and L.~Liquori.
\newblock A {L}ambda {C}alculus of {I}ncomplete {O}bjects.
\newblock In {\em Proc. of MFCS}, volume 1113, pages 218--229, 1996.

\bibitem[BCC{\etalchar{+}}96]{BruEtAl96}
K.~Bruce, L.~Cardelli, G.~Castagna, The~Hopkins~Object Group, G.~Leavens, and
  B.~Pierce.
\newblock On {B}inary {M}ethods.
\newblock {\em Theory and Practice of Object Systems}, 1(3), 1996.

\bibitem[BF98]{BF-98}
Viviana Bono and Kathleen Fisher.
\newblock An imperative, first-order calculus with object extension.
\newblock In {\em Proc. of ECOOP}, volume 1445 of {\em Lecture Notes in Computer Science}, pages 462--497. Springer, 1998.

\bibitem[BL95]{BL-95}
V.~Bono and L.~Liquori.
\newblock A {S}ubtyping for the {F}isher-{H}onsell-{M}itchell {L}ambda
  {C}alculus of {O}bjects.
\newblock In {\em Proc. of CSL}, volume 933, pages 16--30, 1995.

\bibitem[BPF97]{BPF-97}
K.~Bruce, L.~Petersen, and A.~Fiech.
\newblock Subtyping {I}s {N}ot a {G}ood ``{M}atch'' for {O}bject-{O}riented
  {L}anguages.
\newblock In {\em Proc. of ECOOP}, volume 1241, pages 104--127, 1997.

\bibitem[Bru94]{Bruce-94}
K.B. Bruce.
\newblock {A} {P}aradigmatic {O}bject--{O}riented {P}rogramming {L}anguage:
  {D}esign, {S}tatic {T}yping and {S}emantics.
\newblock {\em Journal of Functional Programming}, 4(2):127--206, 1994.

\bibitem[Car95]{Cardelli-95}
L.~Cardelli.
\newblock A {L}anguage with {D}istributed {S}cope.
\newblock {\em Computing System}, 8(1):27--59, 1995.

\bibitem[Cas95]{Castagna-95}
G.~Castagna.
\newblock Covariance and contravariance: conflict without a cause.
\newblock {\em ACM Transactions on Programming Languages and Systems},
  17(3):431--447, 1995.

\bibitem[Cas96]{Castagna-book}
G.~Castagna.
\newblock {\em Object-Oriented Programming: A Unified Foundation}.
\newblock Progress in Theoretical Computer Science. Birk\"auser, Boston, 1996.

\bibitem[CG09]{CohenG09}
Tal Cohen and Joseph Gil.
\newblock Three approaches to object evolution.
\newblock In {\em Proc. of PPPJ}, pages 57--66. {ACM}, 2009.

\bibitem[CHJ12]{ChughHJ12}
Ravi Chugh, David Herman, and Ranjit Jhala.
\newblock Dependent types for javascript.
\newblock {\em SIGPLAN Not.}, 47(10):587--606, 2012.

\bibitem[DDDG01]{fickle1:fool01}
Sophia Drossopoulou, Ferruccio Damiani, Mariangiola Dezani{-}Ciancaglini, and
  Paola Giannini.
\newblock Fickle: Dynamic object re-classification.
\newblock In {\em Proc. of ECOOP} 2001,
volume 2072 of {\em Lecture Notes in Computer Science}, pages 130--149. Springer, 2001.

\bibitem[DDDG02]{fickle2:toplas02}
Sophia Drossopoulou, Ferruccio Damiani, Mariangiola Dezani{-}Ciancaglini, and
  Paola Giannini.
\newblock More dynamic object reclassification:
  Fickle\({}_{\mbox{{\(\vert\)}{\(\vert\)}}}\).
\newblock {\em {ACM} Trans. Program. Lang. Syst.}, 24(2):153--191, 2002.

\bibitem[DDG03]{fickle3:ictcs03}
Ferruccio Damiani, Sophia Drossopoulou, and Paola Giannini.
\newblock Refined effects for unanticipated object re-classification:
  Fickle\({}_{\mbox{3}}\).
\newblock In {\em Proc. of ICTCS},
volume 2841 of {\em Lecture Notes in Computer Science}, pages 97--110. Springer, 2003.

\bibitem[DGHL98]{DBLP:conf/oopsla/GianantonioHL98}
Pietro Di~Gianantonio, Furio Honsell, and Luigi Liquori.
\newblock A lambda calculus of objects with self-inflicted extension.
\newblock In {\em Proc. of OOPSLA}, pages 166--178. {ACM}, 1998.

\bibitem[FHM94]{FHM-94}
K.~Fisher, F.~Honsell, and J.~C. Mitchell.
\newblock A {L}ambda {C}alculus of {O}bjects and {M}ethod {S}pecialization.
\newblock {\em Nordic Journal of Computing}, 1(1):3--37, 1994.

\bibitem[FM94]{FM-94}
K.~Fisher and J.~C. Michell.
\newblock Notes on {T}yped {O}bject-{O}riented {P}rogramming.
\newblock In {\em Proc. of TACS}, volume 789, pages 844--885, 1994.

\bibitem[FM95]{FM-95}
K.~Fisher and J.~C. Mitchell.
\newblock A {D}elegation-based {O}bject {C}alculus with {S}ubtyping.
\newblock In {\em Proc. of FCT}, volume 965, pages 42--61, 1995.

\bibitem[FM98]{FM-98}
K.~Fisher and J.~C. Mitchell.
\newblock On the relationship between classes, objects, and data abstraction.
\newblock {\em Theory and Practice of Object Systems}, 1998.
\newblock To appear.

\bibitem[Ghe02]{Ghelli02}
Giorgio Ghelli.
\newblock Foundations for extensible objects with roles.
\newblock {\em Information and Computation}, 175(1):50--75, 2002.

\bibitem[KT10]{KaminaT10}
T~Kamina and T~Tamai.
\newblock A smooth combination of role-based language and context activation.
\newblock In {\em Proceedings of the Ninth Workshop on Foundation of
  Aspect-Oriented Languages (FOAL 2010),}, pages 15--24, 2010.

\bibitem[Liq97]{Liquori-Ecoop-97}
L.~Liquori.
\newblock An {E}xtended {T}heory of {P}rimitive {O}bjects: {F}irst {O}rder
  {S}ystem.
\newblock In {\em Proc. of ECOOP}, volume 1241, pages 146--169, 1997.

\bibitem[MT08]{EpsilonJ}
Supasit Monpratarnchai and Tetsuo Tamai.
\newblock The implementation and execution framework of a role model based
  language, epsilonj.
\newblock In {\em Proc. of SNPD}, pages 269--276, 2008.

\bibitem[R{\'e}m98]{Remy-98}
D.~R{\'e}my.
\newblock From classes to objects via subtyping.
\newblock In {\em Proc. of European Symposium on Programming}, volume 1381 of
  {\em lncs}. springer, 1998.

\bibitem[RGN{\etalchar{+}}14]{RessiaGNPR14}
Jorge Ressia, Tudor G{\^{\i}}rba, Oscar Nierstrasz, Fabrizio Perin, and Lukas
  Renggli.
\newblock Talents: an environment for dynamically composing units of reuse.
\newblock {\em Softw., Pract. Exper.}, 44(4):413--432, 2014.

\bibitem[RS98]{RS-98}
J.G. Riecke and C.~Stone.
\newblock Privacy via {S}ubsumption.
\newblock In {\em Electronic proceedings of FOOL-98}, 1998.

\bibitem[Ser99]{Serrano-ecoop99}
Manuel Serrano.
\newblock Wide classes.
\newblock In {\em Proc. of ECOOP}, volume 1628 of {\em Lecture Notes in Computer Science}, pages 391--415. Springer, 1999.

\bibitem[Tak95]{Takahashi95}
Masako Takahashi.
\newblock Parallel reductions in lambda calculus.
\newblock {\em Inf. Comput.}, 118(1):120--127, April 1995.

\bibitem[Vou01]{Vouillon01}
J{\'e}r\^{o}me Vouillon.
\newblock Combining subsumption and binary methods: An object calculus with
  views.
\newblock In {\em Proc. of POPL} ACM, 2001

\bibitem[Wan87]{Wand-87}
M.~Wand.
\newblock Complete {T}ype {I}nference for {S}imple {O}bjects.
\newblock In {\em Proc. of LICS}, pages 37--44. IEEE Press, 1987.

\bibitem[Zha10]{Zhao10}
Tian Zhao.
\newblock Type inference for scripting languages with implicit extension.
\newblock In {\em ACM SIGPLAN International Workshop on Foundations of
  Object-Oriented Languages}, 2010.

\bibitem[Zha12]{Zhao12}
Tian Zhao.
\newblock Polymorphic type inference for scripting languages with object
  extensions.
\newblock {\em SIGPLAN Not.}, 47(2):37--50, October 2012.

\end{thebibliography}




\section{Typing rules, \ObjP}\label{appendix:plain}


\begin{enumerate}
\itemsep 10pt

\item[] {\bf Well-formed Contexts}

\item[] 
\begin{center}
$
\infer[(Cont{-}\varepsilon)]
 {\varepsilon \vdash ok}
 {}
\quad
\infer[(Cont{-}x)]
 {\Gamma, x \of \sigma \vdash ok}
 {\Gamma \vdash \sigma : \T &
 x \not \in Dom(\Gamma)}
$
\end{center}

\item[] 
\begin{center}
$
\infer[(Cont{-}t)]
 {\Gamma, t \match \proj{R}{m} \vdash ok} 
 {\Gamma \vdash \proj{R}{m} : \T & 
 t \not \in Dom(\Gamma)}
$
\end{center}


\item[] {\bf Well-formed Types}

\item[] 
\begin{center}
$
 \infer[(Type{-}Const)]
 {\Gamma \vdash \iota : \T}
 {\Gamma \vdash ok}
\quad
 \infer[(Type{-}Arrow)]
 {\Gamma \vdash \sigma_1 \arrow \sigma_2 : \T}
 {\Gamma \vdash \sigma_1: \T 
 & 
 \Gamma \vdash \sigma_2 : \T}
$
\end{center}

\item[] 
\begin{center}
$
\infer[(Type{-}Pro_{\l\,\r})]
 {\Gamma \vdash \probject{\,} : \T}
 {\Gamma \vdash ok}
\quad
\infer[(Type{-}Pro)]
 {\Gamma \vdash 
 \probject{R, \m\of\sigma} : \T}
 {\Gamma, t \match \pro{R} \vdash \sigma : \T
 & 
 \m \not \in \meth{R}
 }
$
\end{center}

\item[] 
\begin{center}
$ 
\infer[(Type{-}Extend)]
 {\Gamma \vdash \tau \ovatype \vm{} : \T}
 {\Gamma \vdash \tau \match \pro{R} 
 &
 \vm{} \subseteq \meth{R}
 }
$
\end{center}

\item[] {\bf Matching Rules}

\item[]
\begin{center}
$
\infer[(Match{-}t)]
 {\Gamma \vdash t \ovatype \vm{} \match t \ovatype \vn{}} 
 {\Gamma \vdash t \ovatype \vm{} : \T 
 &
 \vn{} \subseteq \vm{}
 } 
\quad
%
%
\infer[(Match{-}Var)] 
 {\Gamma_1, t \match \tau_1, \Gamma_2 \vdash t \ovatype \vm{}
 \match \tau_2} 
 {\Gamma_1, t \match \tau_1, \Gamma_2 \vdash \tau_1 \ovatype \vm{} \match \tau_2}
 $
\end{center}

\item[]
\begin{center}
$
\infer[(Match{-}Pro)]
 {\Gamma \vdash \pro R_1 \ovatype \vm{}
 \match \pro R_2 \ovatype \vn{}}
 {\Gamma \vdash \pro R_1 \ovatype \vm{}: \T
 & 
 \Gamma \vdash \pro R_2 \ovatype \vn{} : \T
 & 
 R_2 \subseteq R_1
 & 
 \vn{} \subseteq \vm{}
 }
$
\end{center}

\item[] {\bf Type Rules for $\lambda$-terms}

\item[] 
\begin{center}
$
\infer[(Const)]
 {\Gamma \vdash c : \iota}
 {\Gamma \vdash ok 
 }
\quad
\infer[(Var)]
 {\Gamma_1, x \of \sigma,\Gamma_2 \vdash x : \sigma}
 {\Gamma_1, x \of \sigma,\Gamma_2 \vdash ok}
$
\end{center}


\item[] 
\begin{center}
$
 \infer[(Abs)]
 {\Gamma \vdash \lambda x. e : \sigma_1 \arrow \sigma_2}
 {\Gamma, x\of\sigma_1 \vdash e:\sigma_2}
\quad
%
 \infer[(Appl)]
 {\Gamma \vdash e_1 e_2: \sigma_2}
 {\Gamma \vdash e_1 : \sigma_1 \arrow \sigma_2
 & 
 \Gamma \vdash e_2 : \sigma_1
 }
$
\end{center}

\item[] {\bf Type Rules for Object Terms}

\item[]
\begin{center}
$
\infer[(Empty)]
 {\Gamma \vdash \l\,\rangle: \probject{\,}}
 {\Gamma \vdash ok}
\quad
%
\infer[(Send)]
 {\Gamma \vdash e \call \n: \sigma[\tau/t]}
 {\Gamma \vdash e : \tau
 & 
 \Gamma \vdash \tau \match \probject{R,\n\of\sigma} \ovatype \vm{},\n
 }
$
\end{center}

\item[]
\begin{center}
$
\infer[(Pre{-}Extend)]
 {\Gamma \vdash e: \probject{R_1, R_2} \ovatype \vm{}}
 {\Gamma \vdash e : \pro{R_1} \ovatype \vm{}
 &
 \Gamma \vdash \probject{R_1, R_2} \ovatype \vm{}: \T
 }
$
\end{center}

\item[]
\begin{center}
$
\infer[(Extend)]
 {\Gamma \vdash 
 \langle e_1 \ova \n = e_2 \rangle : \tau \ovatype \n}
 {\begin{array}{l}
 \Gamma \vdash e_1 : \tau \qquad 
 \Gamma \vdash \tau \match 
 \probject{R, \n\of\sigma} \ovatype \vm{} \\[1mm]
 \Gamma, t \match \probject{R ,\n\of\sigma} \ovatype \vm{},\n \vdash
 \e_2 : t \arrow \sigma
 \end{array}
 }
$
\end{center}

\item[]
\begin{center}
$
\infer[(Override)]
 {\Gamma \vdash 
 \langle e_1 \ova \n = e_2 \rangle: \tau}
 {\begin{array}{l}
 \Gamma \vdash e_1 : \tau \qquad 
 \Gamma \vdash \tau \match 
 \probject{R, \n\of\sigma} \ovatype \vm{},\n \\[1mm]
 \Gamma, t \match \probject{R ,\n\of\sigma} \ovatype \vm{},\n \vdash
 \e_2 : t \arrow \sigma
 \end{array}
 }
$
\end{center}

\item[]
\begin{center}
$
 \infer[(Select)]
 {\Gamma \vdash Sel(\e_1,\n,\e_2) : \sigma[\tau \ovatype \vn{}/t]}
 {\begin{array}{l}
 \Gamma \vdash \e_1 : \tau \qquad 
 \Gamma \vdash \tau \match 
 \probject{R,\n\of\sigma} \ovatype \vm{},\n \\[1mm]
 \Gamma, t \match \probject{R,\n\of\sigma} \ovatype \vm{},\n \vdash 
 \e_2 : t \arrow t \ovatype \vn{}
 \end{array}
 }
$
\end{center} 
 
\end{enumerate}



\section{Extra rules for Subsumption, \ObjPS}\label{appendix:sub}
 



\begin{enumerate}
\itemsep 10pt

\item[] {\bf Extra Well-formed Contexts}

\item[]
\begin{center}
$
\infer[(Cont{-}Obj)]
 {\Gamma, t \match \objj{R}{m} \vdash ok} 
 {\Gamma \vdash \objj{R}{m} : \T & 
 t \not \in Dom(\Gamma)}
$
\end{center}

\item[] {\bf Extra Well-formed Types}

\item[]
\begin{center}
$
 \infer[(Type{-}Obj)]
 {\Gamma \vdash \objj{R}{m} : \T}
 {\Gamma \vdash \proj{R}{m}: \T}
\quad
\infer[(Type{-}Extend{-}Obj)]
 {\Gamma \vdash \tau \ovatype \vm{} : \T}
 {\Gamma \vdash \tau \match \obj{R} 
 & 
 \vm{} \subseteq \meth{R}
 }
$
\end{center}


\item[] {\bf Rules for Rigid Types}

\item[] 
\begin{center}
$
 \infer[(Type{-}Const{-}Rgd)]
 {\Gamma \vdash \iota : \TRIG}
 {\Gamma \vdash ok}
\quad
%
%
%
 \infer[(Type{-}Arrow{-}Rgd)]
 {\Gamma \vdash \sigma_1 \arrow \sigma_2 : \TRIG}
{\Gamma \vdash \sigma_1 : \T & \Gamma \vdash \sigma_2 : \TRIG}
$
\end{center}


\item[]
\begin{center}
$
 \infer[(Type{-}Var{-}Obj)]
 {\Gamma_1, t \match \objj{R}{m},\Gamma_2 \vdash t \ovatype \vn{} : \TRIG}
 {\Gamma_1, t \match \objj{R}{m},\Gamma_2 \vdash t \ovatype \vn{} : \T
 & t ~\mbox{covariant in}~ R }
$
\end{center}


%
%


\item[]
\begin{center}
$
 \infer [(Type{-}Obj{-}Rdg)]
 {\Gamma \vdash \objj{\l \vm{k}:\vs{k} \r}{n} : \TRIG}
 {\Gamma \vdash \objj{\l \vm{k}:\vs{k} \r}{n} : \T
 & \forall i \leq k. \ \Gamma \vdash \sigma_i : \TRIG
 \land t ~\mbox{covariant in}~\sigma_i}
$
\end{center}



\item[] {\bf Extra Matching Rules}


\item[]
\begin{center}
$
 \infer[(Match{-}Arrow)]
 {\Gamma \vdash \sigma_1 \arrow \sigma_2 \match \sigma_1' \arrow \sigma_2'}
 {\Gamma \vdash \sigma_1' \match \sigma_1
 & 
 \Gamma \vdash \sigma_2 \match \sigma_2'
 & 
 \Gamma \vdash \sigma_1 : \TRIG}
$
\end{center}

 
\item[]
\begin{center}
$
\infer[(Promote)]
 {\Gamma \vdash \proj{R_1}{m}
 \match \objj{R_2}{n}}
 {
 \Gamma \vdash \proj{R_1}{m}: \T
 & 
 \Gamma \vdash \proj{R_2}{n} : \T
 & 
 R_2 \subseteq R_1
 & 
 \vn{} \subseteq \vm{}
 }
$
\end{center}

\item[]
\begin{center}
$
 \infer[(Match{-}Obj)]
 {\Gamma \vdash \obj{R_1} \ovatype \vm{}
 \match \obj{R_2} \ovatype \vn{}}
 {\Gamma \vdash \pro{R_1} \ovatype \vm{}: \T
 & 
 \Gamma \vdash \pro{R_2} \ovatype \vn{} : \T
 & 
 R_2 \subseteq R_1
 & 
 \vn{} \subseteq \vm{}
 }
$
\end{center}

\item[] {\bf Extra Type Rules for Terms}

\item[]
\begin{center}
$
 \infer[(Extend{-}Obj)]
 {\Gamma \vdash 
 \langle e_1 \ova \n = e_2 \rangle : \tau \ovatype \n}
 {\begin{array}{l}
 \Gamma \vdash e_1 : \tau \qquad 
 \Gamma \vdash \tau \match 
 \object{R, \n\of\sigma} \ovatype \vm{} \\[1mm]
 \Gamma, t \match \object{R ,\n\of\sigma} \ovatype \vm{},\n \vdash
 \e_2 : t \arrow \sigma
 \end{array}
 }
$
\end{center}

\item[]
\begin{center}
$
\infer[(Override{-}Obj)]
 {\Gamma \vdash 
 \langle e_1 \ova \n = e_2 \rangle: \tau}
 {\begin{array}{l}
 \Gamma \vdash e_1 : \tau \qquad 
 \Gamma \vdash \tau \match 
 \object{R, \n\of\sigma} \ovatype \vm{},\n \\[1mm]
 \Gamma, t \match \object{R ,\n\of\sigma} \ovatype \vm{},\n \vdash
 \e_2 : t \arrow \sigma
 \end{array}
 }
$
\end{center}

\item[] 
\begin{center}
$
\infer[(Send{-}Obj)]
 {\Gamma \vdash e \call \n: \sigma[\tau/t]}
 {\Gamma \vdash e : \tau
 & 
 \Gamma \vdash \tau \match \object{R,\n\of\sigma} \ovatype \vm{},\n
 }
$
\end{center}

\item[]
\begin{center}
$
 \infer[(Select{-}Obj)]
 {\Gamma \vdash Sel(\e_1,\n,\e_2) : \sigma[\tau \ovatype \vn{}/t]}
 {\begin{array}{l}
 \Gamma \vdash \e_1 : \tau \qquad 
 \Gamma \vdash \tau \match 
 \object{R,\n\of\sigma} \ovatype \vm{},\n \\[1mm]
 \Gamma, t \match \object{R,\n\of\sigma} \ovatype \vm{},\n \vdash 
 \e_2 : t \arrow t \ovatype \vn{}
 \end{array}
 }
$
\end{center}
 

\item[]
\begin{center}
$
\infer[(Subsume)]
 {\Gamma \vdash \e :\sigma_2}
 {\Gamma \vdash \e :\sigma_1
 & 
 \Gamma \vdash \sigma_1 \match \sigma_2
 & 
 \Gamma \vdash \sigma_2 : \TRIG}
 $
\end{center}


\end{enumerate} 


\newpage
\section{Soundness of the Type System \ObjP}\label{soundness}


\begin{lemma} (Sub-derivation) \label{sub}
  \begin{itemize}
  \item[(i)] If $\Delta$ is a derivation of $\Gamma_1, \Gamma_2 \vdash
    \mathcal{A}$, then there exists a sub-derivation $\Delta' \subseteq \Delta$
    of $\Gamma_1 \vdash ok$.
  \item [(ii)] If $\Delta$ is a derivation of $\Gamma_1,
    x\of\sigma, \Gamma_2 \vdash \mathcal{A}$, then there exists a sub-derivation
    $\Delta' \subseteq \Delta$ of $\Gamma_1 \vdash \sigma: \T$.
  \item [(iii)] If $\Delta$ is a derivation of $\Gamma_1, t
    \match \tau, \Gamma_2 \vdash \mathcal{A}$, then there exists a sub-derivation
    $\Delta' \subseteq \Delta$ of $\Gamma_1 \vdash \tau : \T$.
  \end{itemize}
\end{lemma}  
The three points are proved, separately, by structural induction on the derivation $\Delta$.

\textnormal{(i)} The only cases where the inductive hypothesis cannot
be applied are the cases where the last rule in $\Delta$ is a context
rule (that is, the only kind of rule that can increase the context)
and $\Gamma_2$ is empty. In these cases the thesis coincides with the
hypothesis.  In all the other cases the thesis follows immediately by
an application of the inductive hypothesis.

\textnormal{(ii)} As in point \textnormal{(i)}, either we conclude
immediately by inductive hypothesis or it is the case that
$\Gamma_2$ is empty and the last rule in $\Delta$ is a context rule. In
this latter case the last rule in $\Delta$ is necessarily a $(Cont{-}x)$ rule
deriving $\Gamma_1, x\of\sigma \vdash ok$, and the first
premise of this rule coincides with the thesis.

\textnormal{(iii)} The proof works similarly to point \textnormal{(ii)}.  \qed

\begin{lemma} (Weakening) \label{weak}
    \begin{itemize}
    \item[(i)] If $\Gamma_1, \Gamma_2 \vdash \mathcal{A}$ and $\Gamma_1, \mathcal{C}, \Gamma_2
      \vdash ok$, then $\Gamma_1, \mathcal{C}, \Gamma_2 \vdash \mathcal{A}$.
    \item[(ii)] If $\Gamma_1 \vdash \mathcal{A}$ and $\Gamma_1, \Gamma_2 \vdash ok$,
      then $\Gamma_1, \Gamma_2 \vdash \mathcal{A}$.
  \end{itemize}
\end{lemma}  
\textnormal{(i)} By structural induction on the derivation $\Delta$ of
$\Gamma_1, \Gamma_2 \vdash \mathcal{A}$.  If the last rule in $\Delta$ has the
context in the conclusion identical to the context in the premise(s),
then it is possible to apply the inductive hypothesis, thus deriving
almost immediately the goal.
In the other cases, if the last rule in $\Delta$ is a $(Cont{-}x)$ or $(Cont{-}t)$ rule, then the proof is trivial, since the second hypothesis coincides with the thesis.  The
remaining cases concern the $(Type{-}Pro)$, $(Abs)$, $(Extend)$
and $(Override)$ rules, which require a more careful treatment.
We detail here only the proof for $(Type{-}Pro)$, since the other rules are handled in a similar way.

In the $(Type{-}Pro)$ case, the hypothesis 
$\Gamma_1, \Gamma_2 \vdash \probject{R,\m\of\sigma} : \T$ follows from:
\begin{equation} \label{weak.1}
\Gamma_1, \Gamma_2, t \match \pro{R} \vdash \sigma : \T
\end{equation}
Let us briefly remark that if the statement $\mathcal{C}$ of the second
hypothesis is equal to $t \match \tau$, for some type $\tau$, then it
is convenient to $\alpha$-convert the type $\probject{R,
\m\of\sigma}$ to avoid clash of variables.  In any case, 
by Lemma~\ref{sub}.(iii) (Sub-derivation), there exists a sub-derivation
of $\Delta$ deriving $\Gamma_1, \Gamma_2 \vdash \pro{R} : \T$, from
which, by inductive hypothesis, $\Gamma_1, \mathcal{C}, \Gamma_2 \vdash
\pro{R} : \T$ and in turn, via the $(Cont{-}t)$ rule, $\Gamma_1, \mathcal{C},
\Gamma_2, t \match \pro{R} \vdash ok$. By using (\ref{weak.1}) and the
inductive hypothesis, we deduce $\Gamma_1, \mathcal{C}, \Gamma_2, t \match \pro{R
} \vdash \sigma : \T$.  Finally we have the thesis via the $(Type{-}Pro)$ rule.

\textnormal{(ii)} By induction on the length of $\Gamma_2$; the proof
uses the previous point (i) and Lemma~\ref{sub}.(i) (Sub-derivation).
\qed


\begin{lemma} (Well-formed object-types) \label{type} 
  \begin{itemize}  
    \item[(i)] $\Gamma \vdash \proj{R}{m}:\T$ if and only
    if $\Gamma \vdash \pro{R}:\T$ and $\vm{} \subseteq \meth{R}$.
    
    \item[(ii)] $\Gamma \vdash t \ovatype \vm{}:\T$ if and only if  $\Gamma$
    contains $t \match \proj{R}{n}$, with $\vm{} \subseteq \meth{R}$.
\end{itemize}
\end{lemma}   
Point \textnormal{(i)} is immediately proved by inspection on the rules
for well-formed types and matching.  Point \textnormal{(ii)} is
proved by inspection on the rules for well-formed contexts, well-formed
types and matching.  \qed

\bigskip Notice that in the following proofs often we will not refer
explicitly to the previous  lemmas, thus considering obvious their application.
%

\begin{proposition} (Matching is well-formed) \label{match-well}

\medskip
If $\Gamma \vdash \tau_1 \match \tau_2$, then
  $\Gamma \vdash \tau_1 : \T$ and $\Gamma \vdash \tau_2 : \T$.
\end{proposition}   
By structural induction on the derivation $\Delta$ of $\Gamma
\vdash \tau_1 \match \tau_2$.
The premises of the $(Match{-}Pro)$ rule coincide with the thesis.  If the last rule in $\Delta$ is $(Match{-}t)$, we conclude by using its premises and
Lemma~\ref{type}.(ii) (Well-formed object-types).
If the last rule in $\Delta$ is $(Match{-}Var)$, then the judgment $\Gamma_1, t
\match \rho, \Gamma_2 \vdash t \ovatype \vm{} \match \tau_2$ is derived from
$\Gamma_1, t \match \rho, \Gamma_2 \vdash \rho \ovatype \vm{}
\match \tau_2$.  By inductive hypothesis $\tau_2$ is well-formed and
$\Gamma_1, t \match \rho, \Gamma_2 \vdash \rho \ovatype \vm{} : \T$.
By inspecting the $(Cont{-}t)$ rule, $\rho$ must be in the form
$\proj{R}{n}$, and by Lemma~\ref{type}.(i) (Well-formed
types) it holds $\vm{} \subseteq \meth{R}$. We can now conclude $\Gamma_1, t \match \rho, \Gamma_2 \vdash t \ovatype \vm{} : \T$ via Lemma~\ref{type}.(ii) (Well-formed object-types). \qed

\begin{lemma} (Matching) \label{matching} 
  \begin{itemize}  
  \item[(i)] $\Gamma \vdash \proj{R_1}{m} \match
    \tau_2$ if and only if $\Gamma \vdash \proj{R_1}{m}:\T$
    and $\Gamma \vdash \tau_2 : \T$ and $\tau_2 \equiv \proj{R_2}{n}$, 
    with $R_2 \subseteq R_1$ and $\vn{} \subseteq \vm{}$.

  \item[(ii)] $\Gamma \vdash \tau_1 \match t \ovatype \vn{}$ if and only if
    $\Gamma \vdash \tau_1 : \T$ and $\tau_1 \equiv t \ovatype \vm{}$, with
    $\vn{} \subseteq \vm{}$.

  \item[(iii)] $\Gamma \vdash t \ovatype \vm{} \match \proj{R_2}{n}$ if and only if $\Gamma$ contains $t \match \proj{R_1}{p}$, with $R_2 \subseteq R_1$ and $\vn{} \subseteq \vm{} \cup \vp{}$.

  \item[(iv)] (Reflexivity) If $\Gamma \vdash \rho : \T$ then $\Gamma
    \vdash \rho \match \rho$.

  \item[(v)] (Transitivity) If $\Gamma \vdash \tau_1 \match \rho$ and
    $\Gamma \vdash \rho \match \tau_2$, then $\Gamma \vdash \tau_1
    \match \tau_2$.

  \item[(vi)] (Uniqueness) If $\Gamma \vdash \tau_1 \match \probject{R_1, m\of\sigma_1}$ and $\Gamma \vdash \tau_1 \match \probject{R_2, m\of\sigma_2}$, then $\sigma_1 \equiv \sigma_2$.

  \item[(vii)] If $\Gamma \vdash \tau_1 \match \tau_2$ and 
      $\Gamma \vdash \tau_2 \tr \m  : \T$, then $\Gamma \vdash \tau_1 \tr \m
    \match \tau_2 \tr\m$.

  \item [(viii)] If $\Gamma \vdash \tau_1 \tr \m \match \proj{R}{n}$,
  then $\Gamma \vdash \tau_1 \match \proj{R}{n} {-} \m$.

\item [(ix)] If $\Gamma \vdash \rho \tr \m : \T$, then $\Gamma \vdash
  \rho \tr \m \match \rho$.
 \end{itemize}
\end{lemma}   
\textnormal{(i)} \textnormal{(ii)} \textnormal{(iii)} The thesis is immediate by inspection on the matching rules.

\textnormal{(iv)} By cases on the form of the object-type $\rho$.
The thesis can be derived immediately using either the $(Match{-}Pro)$ rule or the
$(Match{-}t)$ one.

\textnormal{(v)} By cases on the forms of $\tau_1, \tau_2, \rho$,
using the points \textnormal{(i)}, \textnormal{(ii)}, \textnormal{(iii)}
above.  If $\tau_1 \equiv \proj{R}{m}$, we conclude by a
triple application of point \textnormal{(i)}.  If $\tau_2 \equiv t
\ovatype \vn{}$, we conclude by three applications of point
\textnormal{(ii)}.  If $\tau_1 \equiv t \ovatype \vm{}$ and $\tau_2 \equiv
\proj{R}{n}$, we conclude by reasoning on the form of $\rho$,
using all the points \textnormal{(i)}, \textnormal{(ii)}, \textnormal{(iii)}.

\textnormal{(vi)} By cases on the form of $\rho$, using either point
\textnormal{(i)} or point \textnormal{(iii)}.

\textnormal{(vii)} By cases on the form of $\tau_1$. If $\tau_1 \equiv
\proj{R}{m}$, we have the thesis by point \textnormal{(i)}
and Lemma~\ref{type}.\textnormal{(i)} (Well-formed object-types). If
$\tau_1 \equiv t \ovatype \vm{}$, we reason by cases on the form of $\tau_2$:
if $\tau_2 \equiv \proj{R}{n}$, then we have the thesis by point
\textnormal{(iii)} and the validity of the thesis for {\tt pro}-types; if $\tau_2 \equiv t \ovatype \vn{}$, then we have the thesis by point \textnormal{(ii)}.

\textnormal{(viii)} By cases on the form of $\tau_1$, using either point
\textnormal{(i)} or point \textnormal{(iii)}.

\textnormal{(ix)} By cases on the form of $\rho$, using either point
\textnormal{(i)} or point \textnormal{(ii)} and
Lemma~\ref{type}.(ii) (Well-formed object-types).
\qed

\begin{lemma} (Match Weakening) \label{match-weak}
\begin{itemize}
\item[(i)] If $\Gamma_1, t \match \rho,\Gamma_2 \vdash \mathcal{A}$ and $\Gamma_1 \vdash \tau \match \rho$, with $\tau$ a {\tt pro}-type, then $\Gamma_1, t
\match \tau, \Gamma_2 \vdash \mathcal{A}$.
  
\item[(ii)] If $\Gamma \vdash \probject{R, \n\of\sigma} \ovatype \vm{} : \T$,
then $\Gamma, t \match \probject{R, \n\of\sigma} \ovatype \vm{} \vdash \sigma
: \T$.
\end{itemize}  
\end{lemma} 
\textnormal{(i)} By structural induction on the derivation $\Delta$ of
$\Gamma_1, t \match \rho, \Gamma_2 \vdash \mathcal{A}$.  

The only case where the inductive hypothesis cannot be applied is
when $\Gamma_2$ is empty and the last rule in $\Delta$ is a
rule increasing the length of the context, i.e. the $(Cont{-}t)$ rule.
In fact, $\Gamma, t \match \rho \vdash ok$ is derived from $t {\notin} Dom(\Gamma)$; on the other hand, from the second hypothesis and Lemma~\ref{match-well} we have also that $\Gamma_1 \vdash \tau : \T$, hence we may derive the thesis using the same $(Cont{-}t)$ rule.

For all the other cases but one the application of the inductive hypothesis
and the derivation of  the thesis is immediate, since the last rule in 
$\Delta$ does not use the hypothesis $t \match \rho$ in the context. 
The only rule that can use this hypothesis is
$(Match{-}Var)$: in such a case
$\Gamma_1, t \match \rho, \Gamma_2 \vdash t \ovatype \vm{} \match \upsilon$ is derived from the premise
$\Gamma_1, t \match \rho, \Gamma_2 \vdash \rho \ovatype \vm{} \match \upsilon$.
By inductive hypothesis, we have 
$\Gamma_1, t \match \tau, \Gamma_2 \vdash \rho \ovatype \vm{} \match \upsilon$.
Moreover, from
$\Gamma_1 \vdash \tau \match \rho$ and the Weakening Lemma \ref{weak}, we derive $\Gamma_1, t \match \tau, \Gamma_2 \vdash \tau \match \rho$, from which, by Lemma~\ref{matching}.\textnormal{(vii)},
$\Gamma_1, t \match \tau, \Gamma_2 \vdash \tau \ovatype \vm{} \match \rho \ovatype\vm{}$. Finally, by transitivity of matching (Lemma ~\ref{matching}.\textnormal{(v)}), we have
$\Gamma_1, t \match \tau, \Gamma_2 \vdash \tau \ovatype \vm{} \match \upsilon$, and by an application of the $(Match{-}Var)$ rule we obtain the thesis.

\textnormal{(ii)} First observe that there exists $R_1 \subseteq R$ such
that $\Gamma, t \match \pro{R_1} \vdash \sigma : \T$.
\\
In fact, by Lemma \ref{type}.\textnormal{(i)} (Well-formed object-types), we have $\Gamma \vdash \probject{R, \n\of\sigma}: \T $, that can only be derived by an
application of the $(Type{-}Pro)$ rule; therefore, we have either our
goal or $\Gamma, t \match \probject{R_2, \n\of\sigma} \vdash \alpha :
\T $ for a suitable $R_2$ such that $R \equiv \langle R_2, p\of\alpha \rangle$.
From Lemma~\ref{sub}.\textnormal{(iii)} (Sub-derivation) follows that
$\Gamma \vdash \probject{R_2, \n \of \sigma}: \T$, hence we may conclude the existence of $R_1$.

Now, from $\Gamma, t \match \pro{R_1} \vdash \sigma : \T$, by using Lemma
\ref{sub}.\textnormal{(iii)} (Sub-derivation), the $(Match{-}Pro)$ rule
and point \textnormal{(i)}, we have the thesis.  \qed

\begin{proposition} (Substitution) \label{subst}
  \begin{itemize}        
  \item [(i)] If $\Gamma_1,x\of\sigma,\Gamma_2 \vdash \mathcal{A}$ and $\Gamma_1
    \vdash \e :\sigma$, then $\Gamma_1,\Gamma_2 \vdash \mathcal{A}[\e/x]$.
    
  \item [(ii)] If $\Gamma_1, t \match \tau, \Gamma_2,\Gamma_3 \vdash \mathcal{A}$ and
    $\Gamma_1,t \match \tau, \Gamma_2 \vdash \rho \match \tau$, then
    $\Gamma_1, t \match \tau, \Gamma_2, \Gamma_3[\rho/t] \vdash
    \mathcal{A}[\rho/t]$.
  
  \item [(iii)] If $\Gamma_1, t \match \tau, \Gamma_2 \vdash \mathcal{A}$ and $\Gamma_1 \vdash \rho \match \tau$, then $\Gamma_1, \Gamma_2[\rho/t] \vdash
    \mathcal{A}[\rho/t]$.    
  \end{itemize}
\end{proposition}
\textnormal{(i)} By induction on the derivation $\Delta$ of
$\Gamma_1,x \of \sigma,\Gamma_2 \vdash \mathcal{A}$. The only situation where the inductive hypothesis cannot be immediately applied is when the last rule in $\Delta$ is $(Cont{-}x)$.
In such a case $\Gamma_1,x \of \sigma \vdash ok$ is derived from $\Gamma_1 \vdash \sigma : \T$, from which, by Lemma~\ref{sub}.\textnormal{(i)} (Sub-derivation), we have the thesis.

All the remaining rules can be easily managed by applying the inductive hypothesis, apart from the case where the last rule in $\Delta$ is $(Var)$ and the variable $x$ coincides with the one dealt with by the rule.
In this case the conclusion $\Gamma_1, x\of\sigma, \Gamma_2 \vdash
x:\sigma$ derives from the premise $\Gamma_1, x\of\sigma, \Gamma_2 \vdash ok$ and so $\Gamma_1,\Gamma_2 \vdash ok$ by induction.
By the second hypothesis $\Gamma_1 \vdash e:\sigma$ and Lemma~\ref{weak} (Weakening), we deduce $\Gamma_1,\Gamma_2 \vdash e:\sigma$.

\textnormal{(ii)} By induction on the derivation $\Delta$ of $\Gamma_1,t
\match \tau,\Gamma_2,\Gamma_3 \vdash \mathcal{A}$.
As in the previous point, the only case where the inductive hypothesis cannot be applied is when the last rule in $\Delta$ is a context rule; in this case the hypothesis coincides with the thesis.

About the remaining rules, the only non-trivial case is when the last
rule in $\Delta$ is $(Match{-}Var)$ (the only rule that can use the
judgment $t \match \tau$ of the context) and the type
variable $t$ coincides with the one dealt with by the rule.
In this case the conclusion $\Gamma_1,t \match \tau,\Gamma_2,\Gamma_3 \vdash t \ovatype \vm{} \match \tau_2$ derives from the premise $\Gamma_1,t \match
\tau,\Gamma_2,\Gamma_3 \vdash \tau \ovatype \vm{} \match \tau_2$; then, by
inductive hypothesis, $\Gamma_1,t \match \tau,\Gamma_2,\Gamma_3[\rho/t]
\vdash (\tau \ovatype \vm{} \match \tau_2)[\rho/t]$.
By the side condition on $(Cont{-}t)$, $t$ cannot be free in $\tau$ and,
by Lemma \ref{matching} (i), neither in $\tau_2$;
hence, the above judgment can be written as $\Gamma_1,t \match
\tau,\Gamma_2,\Gamma_3[\rho/t] \vdash \tau \ovatype \vm{} \match \tau_2$.  
On the other hand, from the second hypothesis $\Gamma_1,t \match
\tau, \Gamma_2 \vdash \rho \match \tau$ we can derive $\Gamma_1,t \match
\tau, \Gamma_2, \Gamma_3[\rho/t] \vdash \rho \ovatype \vm{} \match \tau \ovatype\vm{}$ by Lemma~\ref{weak}.\textnormal{(ii)} (Weakening) and
Lemma~\ref{matching}.\textnormal{(vii)}, and from the transitivity of
matching (Lemma~\ref{matching}.\textnormal{(v)}) we can conclude
$\Gamma_1,t \match \tau, \Gamma_2, \Gamma_3[\rho/t] \vdash \rho \ovatype \vm{} \match \tau_2$.

\textnormal{(iii)} By the previous point we can derive $\Gamma_1, t
\match \tau, \Gamma_2[\rho/t] \vdash \mathcal{A}[\rho/t]$. Now, via an immediate induction, one can prove that if $\Gamma_1, t \match \tau,
\Gamma_2 \vdash \mathcal{A}$ and $t$ is not free in $\Gamma_2$ nor in $\mathcal{A}$, then $\Gamma_1, \Gamma_2 \vdash \mathcal{A}$. The thesis follows immediately from such a property.  \qed

\begin{proposition} (Types of expressions are well-formed) \label{expr}
  
  \medskip
  If $\Gamma \vdash \e :\beta$, then $\Gamma
  \vdash \beta :\T $.
\end{proposition}

By structural induction on the derivation $\Delta$ of $\Gamma
\vdash \e : \beta$.  In this proof we need to consider explicitly all
the possible cases for the last rule in $\Delta$; each case is quite
simple but needs specific arguments.

(Rules for $\lambda$-terms)
If the last rule in $\Delta$ is $(Const)$, we derive the thesis via $(Type{-}Const)$.
To address the $(Var)$ rule we use Lemma~\ref{sub}.\textnormal{(ii)} (Sub-derivation) and Lemma~\ref{weak}.\textnormal{(i)} (Weakening).
For the $(Abs)$ rule one applies the inductive hypothesis, Lemma~\ref{sub}.(ii)
(Sub-derivation), Lemma~\ref{subst}.\textnormal{(i)} (Substitution),
and the $(Type{-}Arrow)$ rule.
About $(Appl)$, the inductive hypothesis allows us to derive $\Gamma \vdash \alpha \arrow \beta : \T$; this judgment can only be derived through the $(Type{-}Arrow)$ rule, whose second premise is precisely the thesis.
  
(Rules for object terms)
%
The thesis is trivial for the $(Empty)$, $(Pre{-}Extend)$ and $(Override)$ rules.
In the $(Extend)$ case, $\Gamma \vdash \langle e_1 \ova \n = e_2\rangle:
\tau \tr \n$ is derived from $\Gamma \vdash \tau \match \probject{R,
\n\of\sigma} \ovatype \vm{}$; by Proposition~\ref{match-well} and Lemma~\ref{type}.\textnormal{(i)}, we have  $\Gamma \vdash \probject{R, \n\of\sigma}\ovatype \vm{},\n : \T$; by Lemma~\ref{matching}.\textnormal{(vii)}, $\Gamma \vdash \tau \tr \n \match \probject{R, \n\of\sigma}\ovatype \vm{},\n$, and so we conclude by Proposition~\ref{match-well}.
The two remaining cases are more complex.

 $(Send)$ We have that 
 $\Gamma \vdash e \call \n: \sigma[\tau/t]$ is derived from
 $\Gamma \vdash \tau \match \probject{R,\n\of\sigma} \ovatype \vm{},\n$, from which, by Proposition~\ref{match-well}, we derive $\Gamma \vdash \probject{R,\n\of\sigma} \ovatype \vm{},\n : \T$ and, in turn, $\Gamma, t \match
  \probject{R,\n\of\sigma} \ovatype \vm{},\n \vdash \sigma : \T$ by
  Lemma~\ref{match-weak}.\textnormal{(ii)}; finally,
  by Proposition~\ref{subst}.\textnormal{(iii)} (Substitution), we can
  conclude that $\Gamma \vdash \sigma[\tau/t] : \T$.

 $(Select)$ We have that
 $\Gamma \vdash Sel(\e_1,\n,\e_2) : \sigma[(\tau \ovatype \vn{})/t]$ 
is derived from both
 $\Gamma, t \match \probject{R,\n\of\sigma} \ovatype \vm{},\n \vdash  \e_2 : t \arrow (t \ovatype \vn{})$ and
 $\Gamma \vdash \tau \match \probject{R,\n\of\sigma} \ovatype \vm{},\n$. 
By inductive hypothesis,
 $\Gamma, t \match \probject{R,\n\of\sigma} \ovatype \vm{},\n \vdash t \arrow (t \ovatype \vn{}) : \T$ and, by Proposition~\ref{subst}.\textnormal{(iii)} (Substitution),
$\Gamma \vdash \tau \arrow (\tau \ovatype \vn{}) : \T$; then, since this latter judgment can only be obtained via the $(Type{-}Arrow)$ rule, we deduce
$\Gamma \vdash \tau \ovatype \vn{} : \T$.
Further, we have 
$\Gamma \vdash \tau \ovatype \vn{} \match \probject{R,\n\of\sigma} \ovatype \vm{},\n$ by case analysis and Lemma~\ref{matching}.(i)-(iii), from which the thesis by Lemma~\ref{match-weak}.(ii) and Proposition~\ref{subst}.(iii) (Substitution).
 \qed



\begin{theorem} (Subject Reduction, \ObjP) \label{sr}
If $\Gamma \vdash \e :\beta$ and $\e \eval \e'$, then $\Gamma \vdash \e':\beta$.
\end{theorem}    
%
We prove that the type is preserved by each of the four
reduction rules $(Beta)$, $(Selection)$, $(Success)$ and $(Next)$.
  
\medskip 
  
$(Beta)$ The derivation $\Delta$ of $\Gamma \vdash (\lambda x. e_1)
e_2 : \beta$ needs to terminate with a rule $(Appl)$, deriving $\Gamma
\vdash (\lambda x. e_1) e_2 : \alpha$, potentially followed by some
applications of $(Pre{-}Extend)$. Let the premises of $(Appl)$ be
$\Gamma \vdash (\lambda x. e_1) : \sigma \arrow \alpha$ and $\Gamma
\vdash e_2 : \sigma$ for a suitable $\sigma$; in turn, the first
judgment has to be derived from $\Gamma, x\of\sigma \vdash e_1 :
\alpha$ via the $(Abs)$ rule. By Proposition~\ref{subst}.\textnormal{(i)}
(Substitution), we conclude $\Gamma \vdash (e_1 : \alpha)[\e_2/x] \equiv
e_1[\e_2/x] : \alpha$; then, by repeating the potential applications of
$(Pre{-}Extend)$ in $\Delta$, we have the thesis.
  
\medskip 

$(Selection)$ The derivation $\Delta$ of $\Gamma \vdash \e \call \n :
\beta$ has to terminate with a $(Send)$ rule, deriving $\Gamma \vdash
\e \call \n : \sigma[\tau/t]$, potentially followed by
applications of $(Pre{-}Extend)$.  The premises of $(Send)$
are $\Gamma \vdash e : \tau$ and $\Gamma \vdash \tau \match
\probject{R,\n\of\sigma} \ovatype \vm{},\n$.  From this latter judgment, by
Lemma \ref{match-well} (Matching is well-formed) and the rules
$(Cont{-}t)$, $(Match{-}Pro)$, $(Match{-}Var)$, $(Type{-}Extend)$, $(Cont{-}x)$,
$(Var)$, and $(Abs)$,
one can derive $\Gamma, t \match \probject{R,\n\of\sigma}\ovatype \vm{},\n
\vdash \lambda s . s : t \arrow t$.  From the above premises, by
applying the $(Select)$ rule, we have $\Gamma \vdash Sel(\e,\n,
\lambda s . s) : \sigma[\tau/t]$ and, by repeating the potential applications
of $(Pre{-}Extend)$ in $\Delta$, the thesis.

\medskip 

$(Success)$ The derivation $\Delta$ of $\Gamma \vdash Sel(\l \e_1 \ova
\n = \e_2 \rangle,\n,\e_3) : \beta$ must terminate with a
$(Select)$ rule, deriving $\Gamma \vdash Sel(\l \e_1 \ova \n = \e_2
\rangle,\n,\e_3) : \sigma[(\tau \ovatype \vn{})/t]$, potentially followed by
applications of $(Pre{-}Extend)$.  The premises of $(Select)$ are:
\begin{equation} \label{su.1}
\Gamma \vdash \l \e_1 \ova \n = \e_2 \rangle : \tau
\end{equation}
\begin{equation} \label{su.2}
\Gamma \vdash \tau \match \probject{R,\n\of\sigma} \ovatype \vm{},\n
\end{equation}
\begin{equation} \label{su.3}
\Gamma, t \match \probject{R,\n\of\sigma} \ovatype \vm{},\n \vdash \e_3 : t \arrow t \ovatype \vn{}
\end{equation}
From (\ref{su.2}) and (\ref{su.3}), through the Substitution Lemma, we have
$\Gamma \vdash \e_3 : \tau \arrow \tau \ovatype \vn{}$; from this latter judgment and 
(\ref{su.1}), by the $(Appl)$ rule, we derive:
\begin{equation} \label{su.5}
\Gamma \vdash \e_3 \l \e_1 \ova \n = \e_2 \rangle : \tau \ovatype \vn{}
\end{equation}
The judgment (\ref{su.1}) can only be obtained using either the $(Extend)$
rule or the $(Override)$ one, potentially followed by some applications of $(Pre{-}Extend)$.
Here we consider only the case where $(Extend)$ is applied, since $(Override)$ can be managed similarly, with the difference that in some points the proof is simpler.  Hence, let us assume that $(Extend)$ derives $\Gamma \vdash \l \e_1 \ova \n = \e_2 \ra : \rho \tr \n $ from the premise $\Gamma \vdash e_{1} : \rho$ and:
\begin{equation} \label{su.7.2}
\Gamma \vdash \rho \match \probject{R_1, \n\of\sigma_1} \ovatype \vp{}
\end{equation}
\begin{equation} \label{su.7.3}
\Gamma, t \match \probject{R_1 ,\n\of\sigma_1} \ovatype \vp{}, \n \vdash
\e_2 :  t \rightarrow \sigma_1
\end{equation}
By inspection of the $(Pre{-}Extend)$ rule, we can readily derive
$\Gamma \vdash \tau \match \rho \tr \n$.
From (\ref{su.7.2}), by Lemma~\ref{matching}.\textnormal{(vii)}, we have $\Gamma \vdash \rho \tr {\n} \match \probject{R_1, \n\of\sigma_1} \ovatype \vp{}, \n $, and, by transitivity of matching,  $\Gamma \vdash \tau \match \probject{R_1, \n\of\sigma_1} \ovatype \vp{}, \n$.
From this latter judgment and (\ref{su.2}), by
Lemma~\ref{matching}.\textnormal{(vi)} (Matching uniqueness), it follows
that $\sigma \equiv \sigma_1$. 

On the other hand, by Lemma~\ref{matching}.(ix), we have 
$
\Gamma \vdash \tau \ovatype \vn{} \match \tau
$ 
and, by transitivity of matching,
$
\Gamma \vdash \tau \ovatype \vn{} \match \probject{R_1, \n\of\sigma} \ovatype \vp{},\n.
$
From this latter judgment and (\ref{su.7.3}), by the Substitution Lemma,
we have
$
\Gamma \vdash \e_2 : \tau \ovatype \vn{} \rightarrow \sigma [(\tau \ovatype \vn{})/t]
$,
and, in turn, from this and (\ref{su.5}), 
$
\Gamma \vdash \e_2 (\e_3 \l \e_1 \ova \n = \e_2 \rangle) : \sigma [(\tau \ovatype \vn{})/t]$ via the $(Appl)$ rule.
Finally, by repeating the potential applications of $(Pre{-}Extend)$ in $\Delta$, we obtain the thesis.

\medskip

$(Next)$  As argued for $(Success)$, the derivation of $\Gamma \vdash Sel(\l \e_1 \ova \n = \e_2 \rangle,\m,\e_3) : \beta$ must end with a $(Select)$ rule, deriving 
$
\Gamma \vdash Sel(\l \e_1 \ova \n = \e_2 \rangle,\m,\e_3) : \sigma[(\tau \ovatype \vm{})/t],
$
potentially followed by applications of $(Pre{-}Extend)$.
The premises of $(Select)$ are:
\begin{equation} \label{next.1}
\Gamma \vdash \l \e_1 \ova \n = \e_2 \rangle : \tau
\end{equation}
\begin{equation} \label{next.2}
\Gamma \vdash  \tau \match \probject{R,\m\of\sigma} \ovatype \vn{}, \m
\end{equation}
\begin{equation} \label{next.3}
\Gamma, t \match \probject{R,\m\of\sigma} \ovatype \vn{}, \m  \vdash \e_3 : t \arrow (t \ovatype \vm{}) 
\end{equation}
The judgment (\ref{next.1}) can only be derived using either
the $(Extend)$ rule or the $(Override)$ one, 
potentially followed by some applications of $(Pre{-}Extend)$.
As carried out in the proof for the $(Success)$ rule, we address here only the case where $(Extend)$ is applied, being the $(Override)$ case similar but simpler.

Since $(Pre{-}Extend)$ has been applied and (\ref{next.1}) holds, $\tau$ must be in the form 
$\probject{R_1,\m\of\sigma, \n\of\sigma_1} \ovatype \vn{}, \m, \n$. Hence, let (\ref{next.1}) be derived through $(Pre{-}Extend)$ from:
\[
\Gamma \vdash \l \e_1 \ova \n = \e_2 \ra : \probject{R_2,\m\of\sigma, \n\of\sigma_1} \ovatype \vn{}, \m, \n
\]
(where $R_2 \subseteq R_1$), which, in turn, is derived via the $(Extend)$ rule from the premises:
\begin{equation} \label{next.5.1}
\Gamma \vdash \e_1 :  \probject{R_2,\m\of\sigma, \n\of\sigma_1} \ovatype \vn{}, \m
\end{equation}
\begin{equation} \label{next.5.2}
\Gamma \vdash \probject{R_2,\m\of\sigma, \n\of\sigma_1} \ovatype \vn{}, \m \match 
\probject{R_3, \n\of\sigma_1} \ovatype \vp{}
\end{equation}
\begin{equation} \label{next.5.3}
\Gamma \vdash  t \match \probject{R_3, \n\of\sigma_1} \ovatype \vp{}, \n \vdash   \e_2 :  t \rightarrow \sigma_1
\end{equation}
Then, let $\rho$ represent the type $\probject{R_1,\m\of\sigma, \n\of\sigma_1} \ovatype \vn{}, \m$, i.e. $\tau \equiv \rho \tr n$.
From the judgment (\ref{next.5.1}), by the $(Pre{-}Extend)$ rule, we can derive:
\begin{equation} \label{next.6}
\Gamma \vdash \e_1 : \rho
\end{equation}
By the $(Match{-}Pro)$ rule, we have 
$\Gamma \vdash \rho \tr n \match \probject{R_2,\m\of\sigma, \n\of\sigma_1} \ovatype \vn{}, \m$ and, from this latter judgment, (\ref{next.5.2}) and (\ref{next.5.3}), by transitivity of matching and the Weakening Lemma, we derive
$\Gamma, t \match \rho \tr \n \vdash   \e_2 :  t \rightarrow \sigma_1$.
From it, by means of the $(Extend)$ rule:
\begin{equation} \label{next.added}
\Gamma, t \match \rho, s \of t  \vdash  \langle s \ova \n = \e_2 \rangle : t \tr \n
\end{equation}
Now, through (\ref{next.2}), the $(Match{-}Var)$ rule, and the transitivity of matching, one can derive $\Gamma, t \match \rho  \vdash t \tr \n \match \probject{R,\m\of\sigma} \ovatype \vn{}, \m$.
From this latter judgment and (\ref{next.3}), by Substitution, we obtain
$\Gamma, t \match \rho \vdash  e_3 : t \tr \n \rightarrow t \tr \n \ovatype \vm{}$,
and, from this judgment and (\ref{next.added}), by the $(Appl)$ and  $(Abs)$ rules, we have:
\[
\Gamma, t \match \rho \vdash
\lambda s. e_3 \langle s \ova \n = \e_2 \rangle : t \rightarrow t \tr \n \ovatype \vm{}
\]
This judgment, together with (\ref{next.6}), allows to apply the $(Select)$ rule, thus deriving:
\[
\Gamma \vdash Sel(\e_1,\m,\lambda s . e_3 \langle s \ova \n = e_2 \rangle) 
 : \sigma[(\rho \tr \n \ovatype \vm{})/t]
\]
Finally, we get the thesis via the usual potential applications of $(Pre{-}Extend)$.
\qed

\section{Soundness of the Type System with Subsumption \ObjPS}\label{soundness2}

\begin{theorem} (Subject Reduction, \ObjPS) \label{sr-full}
If $\Gamma \vdash \e :\beta$ and $\e \eval \e'$, then $\Gamma \vdash \e':\beta$.
\end{theorem}
  
As in Theorem \ref{sr}, we prove that the type is preserved by each of the  reduction rules $(Beta)$, $(Selection)$, $(Success)$ and $(Next)$.
In the present case we have to manage the extra difficulty of potential applications of the $(Subsume)$ rule.
 
\medskip 

$(Beta)$ The derivation of $\Gamma \vdash (\lambda x. e_1) e_2 : \beta$
needs to terminate with a rule $(Appl)$, deriving 
 $\Gamma \vdash (\lambda x. e_1) e_2 : \alpha$, potentially followed by
 some applications of $(Pre{-}Extend)$ and $(Subsume)$.
The premises of $(Appl)$ must be $\Gamma \vdash (\lambda x. e_1) :
\sigma \arrow \alpha$ and $\Gamma \vdash e_2 : \sigma$, where the first judgment has to be derived via $(Abs)$, followed by potential applications of $(Subsume)$.
Let $\Gamma \vdash (\lambda x. e_1) : \sigma_1 \arrow \alpha_1$ be the conclusion of the  $(Abs)$ rule, and:
\begin{equation} \label{beta}
\Gamma, x\of\sigma_1 \vdash e_1 : \alpha_1
\end{equation}
 its premise.  
Since the $(Subsume)$ rule has been applied, we have 
$\Gamma \vdash \sigma_1 \arrow \alpha_1 \match \sigma \arrow \alpha$ and 
$\Gamma \vdash \sigma \arrow \alpha: \TRIG$, therefore
$\Gamma \vdash \sigma \match \sigma_1$ and
$\Gamma \vdash \sigma_1 : \TRIG$ and
$\Gamma \vdash \alpha_1 \match \alpha$, where
$\Gamma \vdash \alpha : \TRIG$.
Using these judgments and (\ref{beta}) it is not difficult to prove,
by structural induction, that $\Gamma, x\of\sigma \vdash e_1 : \alpha_1$.
By Substitution Lemma, we have then $\Gamma \vdash e_1[\e_2/x] : \alpha_1$, and, by the $(Subsume)$ rule, $\Gamma \vdash e_1[\e_2/x] : \alpha$, from which
the thesis.

\medskip 

$(Selection)$ This case works as for the system without subsumption.

\medskip 

$(Success)$ As in Theorem \ref{sr} (type system without subsumption), we can start by asserting that the derivation $\Delta$ of $\Gamma \vdash Sel(\l \e_1 \ova \n = \e_2 \rangle,\n,\e_3) : \beta$ must end with a $(Select)$ rule, deriving $\Gamma \vdash Sel(\l \e_1 \ova \n = \e_2 \rangle,\n,\e_3) : \sigma[(\tau \ovatype \vn{})/t]$.
This is potentially followed by applications of the $(Pre{-}Extend)$ rule and, in the present case, also the $(Subsume)$ rule.
The premises of $(Select)$ are the following:
\begin{equation} \label{su2.1}
\Gamma \vdash \l \e_1 \ova \n = \e_2 \rangle : \tau
\end{equation}
\begin{equation} \label{su2.2}
\Gamma \vdash \tau \match \object{R,\n\of\sigma} \ovatype \vm{},\n
\end{equation}
\begin{equation} \label{su2.3}
\Gamma, t \match \object{R,\n\of\sigma} \ovatype \vm{},\n \vdash \e_3 : t \arrow t \ovatype \vn{}
\end{equation}
If the judgment (\ref{su2.1}) was not obtained by an application of the $(Subsume)$ rule, we could repeat the steps argued to prove Theorem \ref{sr}.
%
In fact, we address here the case where (\ref{su2.1}) is derived by a single application of $(Subsume)$ (it sufficient to consider a single application, because consecutive applications can be always compacted into a single one).
Hence, let the premises of $(Subsume)$ be:
\begin{equation} \label{su2.4}
\Gamma \vdash \l \e_1 \ova \n = \e_2 \rangle : \rho
\end{equation}
\begin{equation} \label{su2.5}
\Gamma \vdash \rho \match \tau
\end{equation}
\begin{equation} \label{su2.6}
\Gamma \vdash \tau : \TRIG
\end{equation}
From the judgments (\ref{su2.2}), (\ref{su2.5}) and (\ref{su2.3}),
by transitivity of matching and Substitution, we have
$\Gamma \vdash \e_2 : \rho \arrow \rho \ovatype \vn{}$. From this and (\ref{su2.4}), by the $(Appl)$ rule, we derive:
\begin{equation} \label{su2.7}
\Gamma \vdash \e_3 \l \e_1 \ova \n = \e_2 \rangle : \rho \ovatype \vn{}
\end{equation}
Again, by repeating the steps carried out for Theorem \ref{sr} (case analysis on the derivation of (\ref{su2.4})), we can prove that
$
\Gamma
\vdash \e_2(\e_3 \l \e_1 \ova \n = \e_2 \rangle) : \sigma [(\rho \ovatype \vn{})/t]
$.
 
Now, from (\ref{su2.2}) and (\ref{su2.6}) follows that $t$ is covariant in $\sigma$ and 
$
\Gamma \vdash \sigma : \TRIG
$,
and from Lemma \ref{rigid} that
$
\Gamma \vdash \sigma [(\rho \ovatype \vn{})/t] \match \sigma [(\tau \ovatype \vn{})/t] 
$ 
and
$
\Gamma \vdash  \sigma [(\tau \ovatype \vn{})/t] : \TRIG
$.
Finally, by an application of the $(Subsume)$ rule, we have
$
\Gamma
\vdash \e_2 (\e_3 \l \e_1 \ova \m = \e_2 \rangle) : \sigma [(\tau \ovatype \vn{})/t]
$,
and from this the thesis via the applications of $(Pre{-}Extend)$ potentially in $\Delta$.

\medskip 

$(Next)$ As in the version without subsumption,
we start from the derivation $\Delta$ of $\Gamma \vdash Sel(\l \e_1 \ova \n = \e_2 \rangle,\m,\e_3) : \beta$, which has to terminate with a $(Select)$ rule, deriving 
$
\Gamma \vdash Sel(\l \e_1 \ova \n = \e_2 \rangle,\m,\e) : \sigma[(\tau \ovatype \vm{})/t],
$
potentially followed by applications of the $(Pre{-}Extend)$ and $(Subsume)$ rules.
Let the premises of $(Select)$ be:
\begin{equation} \label{next2.1}
\Gamma \vdash \l \e_1 \ova \n = \e_2 \rangle : \tau
\end{equation}
\begin{equation} \label{next2.2}
\Gamma \vdash \tau \match \object{R,\m\of\sigma} \ovatype \vn{}, \m
\end{equation}
\begin{equation} \label{next2.3}
\Gamma, t \match \object{R,\m\of\sigma} \ovatype \vn{}, \m \vdash \e_3 : t \arrow (t \ovatype \vm{}) 
\end{equation}
If the judgment (\ref{next2.1}) was not obtained by an application of the $(Subsume)$ rule, we could repeat the steps argued to prove Theorem \ref{sr}.
%
%
%
Then, we address here the case where (\ref{next2.1}) is derived by a single application of $(Subsume)$, from the premises:
\begin{equation} \label{next2.4}
\Gamma \vdash \l \e_1 \ova \n = \e_2 \rangle : \rho
\end{equation}
\begin{equation} \label{next2.5}
\Gamma \vdash \rho \match \tau
\end{equation}
\begin{equation} \label{next2.6}
\Gamma \vdash \tau : \TRIG
\end{equation}
From these hypotheses, by repeating the same steps argued for the proof 
without subsumption (case analysis on the derivation of the judgment 
(\ref{next2.4})), we deduce:
\[
\Gamma \vdash Sel(\e_1,\m,\lambda s . e_3 \langle s \ova \n = e_2 \rangle) 
 : \sigma[(\rho \tr \n \ovatype \vm{})/t]
\]
Finally, the proof can be accomplished as in the $(Success)$ case, by applying Lemma \ref{rigid} and by means of the $(Subsume)$ and $(Pre{-}Extend)$ rules. \qed
  


\end{document}